\documentclass[12pt]{elsart}

\usepackage{amssymb}
\usepackage{graphics}
\usepackage{graphicx}
\newcommand{\xx}{\mbox{\boldmath$x$}}

\newcommand{\vv}{\mbox{\boldmath$v$}}

\newcommand{\re}{r_e}
\newcommand{\xxie}{\xi_e}
\newcommand{\Zq}{N_{q}}

\newcommand{\bftau}{\mbox{\boldmath$\tau$}}
\newcommand{\Tphys}{T_{\rm phys}}
\newcommand{\Cv}{C_{\rm\scriptscriptstyle V}}

\newcommand{\ue}{u_e}
\newcommand{\ve}{v_e}
\newcommand{\xie}{\xi_e}
\begin{document}

\begin{frontmatter}


\title{\large Gravothermal Catastrophe and Tsallis' Generalized 
Entropy of Self-Gravitating Systems III. \\
quasi-equilibrium structure using normalized $q$-values
}
\author[taruya]{Atsushi Taruya}
\address[taruya]{Research Center for the Early Universe(RESCEU), 
School of Science, University of Tokyo, Tokyo 113-0033, Japan}
\ead{ataruya@utap.phys.s.u-tokyo.ac.jp}


\author[sakagami]{Masa-aki Sakagami}
\address[sakagami]{Department of Fundamental Sciences, FIHS, 
Kyoto University, Kyoto 606-8501, Japan}
\ead{sakagami@phys.h.kyoto-u.ac.jp}
\begin{abstract}
We revisit the issues on the thermodynamic property of 
stellar self-gravitating system arising from Tsallis' non-extensive 
entropy. Previous papers (Physica A 307 (2002) 185; 
{\it ibid}. (2003) in press (cond-mat/0204315)) have revealed that the 
extremum-state of Tsallis entropy, so-called {\it stellar polytrope}, 
has consistent thermodynamic structure, which predicts 
the thermodynamic instability due to the negative specific heat. However, 
their analyses heavily relies on the old Tsallis formalism using standard 
linear mean values. In this paper, extending our previous study, we focus 
on the quasi-equilibrium structure based on the standard framework by means 
of the normalized $q$-expectation values. It then turns out that the new 
extremum-state of Tsallis entropy essentially remains unchanged 
from the previous result, i.e., the stellar quasi-equilibrium distribution 
can be described by the stellar polytrope. While the thermodynamic 
stability for a system confined in an adiabatic wall completely agrees 
with the previous study and thereby the stability/instability criterion 
remains unchanged, the stability analysis reveals a new 
equilibrium property for the system surrounded by a thermal bath. 
In any case, the stability/instability criteria are consistently 
explained from the presence of negative specific heat and within 
the formalism, the stellar polytrope is characterized as a 
plausible non-extensive meta-equilibrium state. 
\end{abstract}
\end{frontmatter}
%
%
%
%
%
%
%
\section{Introduction}
\label{sec: intro}
%
%
%
%
%
%
%
In many astrophysical problems involving self-gravitating many-body 
system, thermodynamic or statistical mechanical treatment 
usually loses its power due to the peculiar feature of long-range 
attractive force. In some restricted cases with the long-term 
evolution of stellar self-gravitating systems, 
however, thermodynamic stability criterion or statistical mechanical analysis 
recovers the physical relevance and plays an essential role 
in predicting the fate of such system. In fact, it is well-known that 
late-time stellar dynamical evolution of the globular clusters as a 
real astrophysical system is driven by 
so-called {\it gravothermal catastrophe}, i.e., thermodynamic instability 
arising from the negative specific heat, which is widely accepted as a 
fundamental astrophysical process \cite{BT1987,EHI1987,MH1997}. 
Historically, the gravothermal 
catastrophe has been investigated in detail considering the very idealized 
situation, i.e., stellar self-gravitating system consisting of 
many particles confined within a cavity of hard sphere 
\cite{Antonov1962,LW1968,HS1978,Padmanabhan1989,Padmanabhan1990}. 
In particular, special attention to the statistical mechanical approach 
using the Boltzmann-Gibbs entropy has been paid.

Recently,  this classic issue was re-analyzed by present authors 
on the basis of the non-extensive 
thermostatistics with Tsallis' generalized entropy \cite{T1988} 
(for comprehensive review of Tsallis' non-extensive formalism and 
its application to the other subject of physics, see 
Refs.\cite{T1999,AO2001}). 
In contrast to the Boltzmann-Gibbs entropy, the quasi-equilibrium 
distribution\footnote{ Here and in what follows, we use 
the term {\it quasi-equilibrium} rather than the (thermal) equilibrium,  
since there is no strict thermal equilibrium in a self-gravitating 
system. The quasi-equilibrium state means that the system is, at least,   
stable in a dynamical equilibrium state.  
}

characterized by the Tsallis entropy can be reduced to the stellar 
polytropic system\cite{PP1993,PP1999}. Evaluating the second variation of 
entropy around the quasi-equilibrium state, we have developed the stability 
analysis and discussed the criterion for onset of gravothermal 
catastrophe (Ref.\cite{TS2002a}, hereafter paper I). Further, 
to clarify the origin of this instability,  
thermodynamic properties has been investigated in detail 
calculating the specific heat of the stellar polytrope 
(Ref.\cite{TS2002b}, hereafter paper II). 
The most noticeable thing in their papers is that the existence of 
thermodynamic instability indicated from the second variation of 
entropy or free-energy is completely explained from the presence of 
negative specific heat. 
As a consequence, the gravothermal instability appears at the 
polytrope indices $n>5$ for a system confined in an adiabatic wall 
and at $n>3$ for a system surrounded by a thermal bath.

While the results in papers I and II are indeed satisfactory and the 
physical interpretation of the instability is fully consistent with previous 
early works, from a standard viewpoint of the Tsallis non-extensive formalism, 
several important issues still remain unresolved. 
Among these, the most crucial problem is the choice of the 
statistical average in non-extensive thermostatistics. 
The previous analyses have been all investigated utilizing the 
{\it old} Tsallis formalism with the 
standard linear mean values, however, 
a more sophisticated framework by means of 
the normalized $q$-values has been recently presented 
\cite{TMP1998,MNPP2000}. As several authors 
advocated, the analysis using normalized $q$-values is thought to be 
essential, since the undesirable divergences in some physical systems 
can be eliminated safely when introducing the normalized $q$-values. 
Further, non-uniqueness of the Boltzmann-Gibbs theory has been 
shown using the normalized $q$-values \cite{AR2000}. 
Of course, this does not imply that 
all the analyses with standard linear means or un-normalized $q$-values 
lose the physical significance\footnote{Indeed, even the old formalism 
consistently recovers the standard Legendre transform structure leading to 
the usual thermodynamic relations \cite{CT1991,PP1997}.}, 
however, in order to 
pursue the physical reality of the non-extensive thermodynamics,   
thermodynamic structure of the stellar self-gravitating system still 
needs to be investigated and the relation between the old and the new 
formalism must be clarified.

This paper especially focuses on this matter starting from the construction 
of the extremum-entropy state by means of the normalized $q$-values. 
Within a mean-field treatment, we investigate the thermodynamic 
property of the quasi-equilibrium distribution 
surrounded by an adiabatic and a thermally conducting wall. 
In particular, we discuss the existence or the absence of thermodynamic 
instability evaluating the specific heat of the quasi-equilibrium systems. 
Further, in order to check the consistency of the thermodynamic structure,  
the stability/instability criteria obtained from the thermodynamic property 
are re-analyzed from the second variation of entropy and free-energy.   
We found that the extremum state of the Tsallis entropy essentially 
remains unchanged and is described by the stellar polytrope. While 
the stability of the system in an adiabatic wall exactly 
coincides with the results in paper I, the new formalism using 
the normalized $q$-values reveals a new thermodynamic structure 
for a stellar system surrounded by a 
thermal bath. This point will be discussed in detail comparing it with 
the previous results.

The paper is organized as follows. In Section \ref{sec: polytrope}, 
employing the standard Tsallis formalism, 
we revisit the issue on the most probable 
state of stellar quasi-equilibrium distribution on the basis of maximum 
entropy principle. While the resultant extremum state of Tsallis entropy 
reduces to the same stellar polytropic distribution as previously found,  
the new equilibrium distribution has several distinct structures.
Taking fully account of this fact, 
in Section \ref{sec: thermodynamics}, the thermodynamic 
properties of stellar polytropic system are investigated in detail. 
Thermodynamic temperature in stellar system is 
identified through the {\it modified} Clausius relation, 
which is indeed consistent with the recent claim based on the thermodynamic 
zero-th law. Then, we evaluate the specific heat and discuss the existence or 
absence of thermodynamic instability. In Section 
\ref{sec: stability_variation}, the stability/instability criteria are 
re-considered by means of the second variation of entropy and free-energy.  
In contrast to the previous analysis using the old Tsallis formalism, 
the second variation of free-energy shows a distinct thermodynamic 
structure. Nevertheless, 
the zero-eigenvalue problem in each case exactly 
recovers the marginal stability condition inferred from the specific heat. 
Thus, within the new formalism using normalized $q$-values, 
all the analyses are consistent and the stellar polytrope can be 
regarded as a plausible non-extensive thermal state. 
Finally, Section \ref{sec: conclusion} is devoted to the conclusion and 
the discussion.   
%
%
%
%
%
%
%
\section{Maximum entropy principle revisited} 
\label{sec: polytrope}
%
%
%
%
%
%
%
%
Throughout the paper, we pursue to investigate the equilibrium property of 
the stellar self-gravitating system consisting of $N$ particles 
confined in a spherical cavity of radius $\re$. For simplicity, 
each particle has the same mass $m_0$ and interacts via Newton gravity 
only. Then, the total mass becomes $M=Nm_0$. 
In this situation, owing to the maximum entropy principle, we seek 
the most probable quasi-equilibrium distribution 
in an adiabatic treatment. That is, 
we consider the quasi-equilibrium structure as the extremum-entropy 
state in which 
the particles elastically bounce from the wall, keeping the mass $M$ 
and energy $E$ constant.

Following the papers I and II, 
we treat this issue employing the mean-field approach that 
the correlation between particles is smeared out and the system can 
be simply described by the 
one-particle distribution function $f(\xx,\vv)$, defined in six-dimensional 
phase-space $(\xx,\vv)$. In this treatment, the one-particle distribution 
is regarded as a fundamental quantity characterizing the stellar system.  
Let us denote the phase-space element as $h^3=(l_0v_0)^3$ with 
unit length $l_0$ and unit velocity $v_0$ and define the integral measure 
$d^6\bftau\equiv d^3\xx d^3\vv/h^3$. Regarding the function 
$f(\xx,\vv)$ as a fundamental statistical quantity, the energy and 
the mass are respectively expressed as follows: 
\begin{eqnarray}
&&   E= K+U = \int \left\{ \frac{1}{2}\,v^2 + \frac{1}{2}\,
  \Phi(\xx) \right\}\,\,f(\xx,\vv)\,\,d^6\bftau, 
\label{eq: def_E} \\
&&    M = \int \,\, f(\xx,\vv)\,\,d^6\bftau, 
\label{eq: def_M}
\end{eqnarray}
where the quantity $\Phi(\xx)$ is the gravitational potential 
given by 
\begin{eqnarray}
  \Phi(\xx) = -G\int \frac{f(\xx',\vv')}{|\xx-\xx'|}\,d^6\bftau'.
\label{eq: def_Phi}
\end{eqnarray}

On the other hand, in the new framework of Tsallis' non-extensive 
thermostatistics, all the macroscopic observables of the quasi-equilibrium 
system can be characterized by the escort distribution, but the 
escort distribution itself is not thought to be fundamental. 
Rather, there exists a more fundamental probability function 
$p(\xx,\vv)$ that quantifies the phase-space structure. 
With a help of this function, the escort distribution is defined and 
the macroscopic observables are expressed as the normalized 
$q$-expectation value as follows (e.g., Refs.\cite{TMP1998,MNPP2000}): 
\begin{eqnarray}
&  \mbox{escort distribution}~~ &: 
P_q(\xx,\vv)= \frac{\displaystyle \left\{p(\xx,\vv)\right\}^q}
  {\displaystyle \int d^6\bftau \left\{p(\xx,\vv)\right\}^q}\,\,\,, 
\label{eq: def_of_escort}   
\\
&  \mbox{normalized $q$-value}~~ &: 
\langle {\it O}_i\rangle_q = \int d^6 \bftau \,\,{\it O}_i\,\,
P_q(\xx,\vv)\,\,.
\label{eq: def_of_q-values} 
\end{eqnarray}
And based on the fundamental probability $p(\xx,\vv)$, 
the Tsallis entropy is given by
\begin{equation}
        S_q=-\frac{1}{q-1}\,\int d^6\bftau \,
\left[\left\{p(\xx,\vv)\right\}^q-p(\xx,\vv)\right].
\label{eq: Tsallis entropy}
\end{equation}
Note that the probability $p(\xx,\vv)$ satisfies the 
normalization condition:  
\begin{equation}
  \label{eq: normarization_p(x,v)}
  \int\,\,d^6\bftau\,\,p(\xx,\vv) = 1.
\end{equation}

To  apply the above Tsallis formalism to the present problem 
without changing the definition of energy and mass 
(\ref{eq: def_E}) and (\ref{eq: def_M}), we identify the 
one-particle distribution with the escort distribution $P_q$, not 
the probability function $p(\xx,\vv)$:  
\begin{equation}
  \label{eq: escort_dist}
  f(\xx,\vv) = M\,\,\frac{\{p(\xx,\vv)\}^q}{\Zq}\,\,; 
\,\,\,\,\,\,\,\,\,\,\,\,
  \Zq = \int d^6\bftau\,\,\{p(\xx,\vv)\}^q
\end{equation}
so as to satisfy the mass conservation (\ref{eq: def_M}). 
With this identification (\ref{eq: escort_dist}), the normalized 
$q$-values are naturally incorporated into our mean-field treatment. 
As a consequence, later analyses become almost parallel to the 
previous study, in a form-invariant manner.

Now, adopting the relation (\ref{eq: escort_dist}), let us seek the 
extremum-entropy state under the constraints (\ref{eq: def_E}) and 
(\ref{eq: normarization_p(x,v)}). As has been 
discussed recently, there are two approaches that extremize 
the entropy under certain constraints, i.e., the method developed by 
Tsallis, Mendes \& Plastino (TMP)\cite{TMP1998} and the optimal 
Lagrange multiplier(OLM) method by Mart\'inez et al. \cite{MNPP2000}. 
Here, we specifically apply the TMP procedure and find the extremum 
entropy state. The alternative derivation using OLM method is presented 
in Appendix A. The only differences in the final expressions 
between the TMP and the OLM method are the dependence of the quantity 
$\Zq$, which can be summarized in a unified form (see 
Eq.(\ref{eq: extremum_state})  with (\ref{eq: A_Phi0})).

The variational problem in the TMP method is given by the 
following equation: 
\begin{equation}
 \delta\left[ S_q- \alpha\left\{\int d^6\bftau p-1\right\}-
\beta\left\{ \int d^6\bftau 
\left(\frac{1}{2}v^2+\frac{1}{2}\Phi\right)f -  E\right\}\right] = 0,
\label{eq: Max.Ent.Prin}
\end{equation}
where the variables $\alpha$ and $\beta$ denote the Lagrange multipliers. 
The variation with respect to the probability $p(\xx,\vv)$ leads to 
\begin{equation}
\int d^6\bftau\left[-\frac{1}{q-1}\left(q\,p^{q-1}-1\right)\delta p 
-\alpha\,\, \delta p - \beta\, \left(\frac{1}{2}v^2+\Phi\right) 
\delta f\right]= 0. 
\label{eq: Max.Ent.Prin_2}
\end{equation}
Here, we used the fact that 
$\int d^6\bftau \delta\Phi f= \int d^6\bftau \Phi\delta f$. 
The above equation includes the variation $\delta f$, which can be expressed  
with a help of the relation (\ref{eq: escort_dist}): 
\begin{equation}
\delta f(\xx,\vv) = q \,\,f(\xx,\vv) \left\{ \frac{\delta p}{p(\xx,\vv)} 
-\frac{1}{M}\,\int d^6\bftau' f(\xx',\vv') 
\frac{\delta p'(\xx',\vv')}{p(\xx',\vv')} \right\}.
\label{eq: delta_f}
\end{equation}
Then, substituting the above expression into (\ref{eq: Max.Ent.Prin_2}) 
becomes 
\begin{eqnarray}
&& \int d^6\bftau \left[-\frac{1}{q-1}\left(q\,p^{q-1}-1\right)
\right.
\nonumber\\
&&\left. ~~~~~~~~~~~~~~~~~~~~~~~~~~~
-\alpha -\beta\, M\,\,q\,\, \frac{p^{q-1}}{\Zq}\,
\left(\frac{1}{2}v^2+\Phi - \varepsilon \right) \right] \delta p =0.
\label{eq: Max.Ent.Prin_3}
\end{eqnarray}
with the quantity $\varepsilon$ being 
\begin{equation}
\varepsilon = \frac{1}{M}\,\int d^6\bftau 
\left(\frac{1}{2}v^2 + \Phi\right) f. 
\label{eq: epsilon}
\end{equation}
In arriving at the equation (\ref{eq: Max.Ent.Prin_3}), we have exchanged 
the role of the variables $(\xx,\vv)\leftrightarrow(\xx',\vv')$. 
Since the above equation must be satisfied independently of the choice of 
the variation $\delta p$, we obtain
\begin{equation}
- \frac{1}{q-1}  \left( q\, p^{q-1} -1 \right) 
- \alpha -\beta \,\,M \,q\,\frac{p^{q-1}}{\Zq}\,
\left(\frac{1}{2}\,v^2+\Phi-\varepsilon\right)= 0,   
\end{equation}
which reduces to the same power-law distribution 
of $\varepsilon$ as has been derived in papers I and II. That is, 
even in the new formalism, the extremum-entropy state remains unchanged and 
can be described by the so-called {\it stellar polytrope}. Together with 
the result by OLM method (Appendix A), it can be summarized as follows: 
\begin{equation}
f(x,v) = M\,\,\frac{\{p(x,v)\}^q}{\Zq}= 
A \left[\Phi_0 - \frac{1}{2}v^2 - \Phi(x) \right]^{q/(1-q)}, 
\label{eq: extremum_state}
\end{equation}
where we define the constants $A$ and $\Phi_0$: 
\begin{eqnarray}
A = \frac{M}{\Zq}\,\, \left\{ \frac{q(1-q)}{\alpha(1-q)+1} 
        \frac{\beta M }{\Zq'}\right\}^{q/(1-q)},~~~~~~\Phi_0
 = \frac{\Zq'}{\beta M(1-q)} + \varepsilon.
\label{eq: A_Phi0}
\end{eqnarray}
Note that the quantity $\Zq'$ means $\Zq'=\Zq$ for the 
TMP method and $\Zq'=1$ for the result using OLM procedure.

For later analysis, we define the density $\rho(r)$ and the 
isotropic pressure $P(r)$ at the radius $r=|\xx|$ as    
\begin{eqnarray}
\rho(r) &\equiv& \int\frac{d^3\vv}{h^3}\,\,\, f(\xx,\vv),
\nonumber \\
&=& 4\sqrt{2}\pi \,B\left(\frac{3}{2},\frac{1}{1-q}\right)
  \,\frac{A}{h^3}\,\,[\Phi_0-\Phi(r)]^{1/(1-q)+1/2}
\label{eq: density}
\end{eqnarray}
and 
\begin{eqnarray}
P(r) &\equiv& \int\frac{d^3\vv}{h^3} \,\,\,\frac{1}{3}\, v^2 \,
f(\xx,\vv),
\nonumber \\
&=& \frac{8\sqrt{2}\pi }{3}\,B\left(\frac{5}{2},\frac{1}{1-q}\right)
  \,\frac{A}{h^3}\,\,[\Phi_0-\Phi(r)]^{1/(1-q)+3/2}.  
\label{eq: pressure}
\end{eqnarray}
Here, the function $B(a,b)$ denotes the beta function. 
These two equations lead to the following polytropic relation: 
\begin{equation}
  \label{eq: polytrope}
  P(r) = K_n\rho ^{1+1/n}(r), 
\end{equation}
with the polytrope index $n$ given by 
\begin{equation}
  \label{eq: index}
  n = \frac{1}{1-q} \,+ \,\frac{1}{2},
\end{equation}
and with the dimensional constant $K_n$: 
\begin{equation}
  \label{eq: K_n}
  K_n = \frac{1}{n+1}\,
\left[ \frac{4\sqrt{2}\,\pi}{h^3}\,\,
B\left(\frac{3}{2},n-\frac{1}{2}\right)\,\,
\left\{\frac{q(1-q)}{\alpha(1-q)+1}\,\,\frac{\beta\,M}{\Zq'}
\right\}^{n-3/2}\,\frac{M}{\Zq}
\right]^{-1/n}.
\end{equation}
Using these quantities, the one-particle distribution can be 
rewritten as follows: 
\begin{eqnarray}
f(x,v)&=& \frac{1}{4\sqrt{2}\pi\,\,B(3/2,n-1/2)}\,\,\,
\frac{\rho\,\,h^3}{ \{ (n+1)\,K_n\,\rho^{1/n}\}^{3/2}}  
\nonumber \\
&&~~~~~~~~~~~~~~~~~~~~~~~~~~~~~~~~
\times\,\,\,\left\{1-\frac{v^2/2}{(n+1)\,K_n\,\rho^{1/n}}\right\}^{n-3/2},    
  \label{eq: poly_dist}
\end{eqnarray}
which agrees with the previous result 
(see Eq.(16) with the identification $(n+1) K_n = (n-3/2)T$ 
in paper I).

While the resultant form of the quasi-equilibrium distribution 
(\ref{eq: poly_dist}) turns out to be invariant irrespective of the 
choice of the statistical averages, we should be aware of the 
two important differences between the old and the new Tsallis 
formalism, which we shall describe below.

First, the relation between the polytrope 
index $n$ and Tsallis parameter $q$ in the polytropic relation 
(\ref{eq: polytrope}) 
differs from the one obtained previously, but is related to 
it through the {\it duality transformation}, $q\leftrightarrow 1/q$ 
(see Eq.(14) in paper I or Eq.(12) in paper II). This property has been 
first addressed in Ref.\cite{TMP1998} in more general context, 
together with the changes in Lagrangian multiplier $\beta$.  
The duality relation implies that all of the thermodynamic properties 
in the new formalism can also be translated into those obtained in the old 
formalism. As shown in section \ref{sec: thermodynamics} and 
\ref{sec: stability_variation}, this is indeed true in the system confined 
in an adiabatic wall(micro-canonical ensemble case), however,  
the duality of thermodynamic structure cannot hold in the system 
surrounded by a thermal bath(canonical ensemble case).

Second, notice that the quasi-equilibrium distribution (\ref{eq: extremum_state}) 
with (\ref{eq: A_Phi0}) contains the new quantities $\Zq$ and $\varepsilon$,    
which implicitly depend on the distribution function itself. In marked contrast to 
the result in old Tsallis formalism,    
this fact gives rise to the non-trivial thermodynamic relations as follows. 
Using the definitions of density and pressure (\ref{eq: density}) and 
(\ref{eq: pressure}),  the quantity $\varepsilon$ becomes
\begin{eqnarray}
  \varepsilon &=& \frac{1}{M}\,
\left\{\frac{3}{2}\,\int d^3x\,P(x) + \int d^3x\,\rho(x)\,\Phi(x)\,\right\}
\nonumber \\
&=& \frac{1}{M}\,
 \left\{\frac{3}{2}\,\int d^3x\,P(x) 
        - \int d^3x\,\rho(x)\,[\Phi_0-\Phi(x)]\,\right\} + \Phi_0.
\nonumber
\end{eqnarray}
Further using the relation $\Phi_0-\Phi(x)= (n+1)(P/\rho)$ from 
(\ref{eq: density}) and (\ref{eq: pressure}) and substituting the equation 
(\ref{eq: A_Phi0}) into the above expression,  the variable $\varepsilon$ 
is cancelled and the equation reduces to 
\begin{equation}
\frac{\Zq'}{\beta}\,=\,\int d^3x\,P(x).
\label{eq: beta-P_relation}
\end{equation}
As for the dependence of $\Zq$, the normalization 
condition (\ref{eq: normarization_p(x,v)}) implies that
\begin{eqnarray}
  \Zq\, = \,\left[ 
\int d^6\bftau\,\,\left\{\,f(\xx,\vv)\,\right\}^{1/q}\,\,
\right]^{-q}.
\nonumber
\end{eqnarray}
Substituting the distribution function (\ref{eq: poly_dist}) into the above 
equation and integrating over the velocity space, after some manipulation, 
one obtains 
\begin{equation}
  \Zq = \,\,c_n\,\,K_n^{(3/2)/(n-1/2)}\,\,
\left\{\int d^3x\,\,\rho^{1+1/n}(x)\,\,\right\}^{-(n-3/2)/(n-1/2)}
  \label{eq: Zq-rho_relation}
\end{equation}
with the constant $c_n$ given by 
\begin{equation}
c_n \,\,= \,\, \frac{4\sqrt{2}\,\pi\,B(3/2,n-1/2)}
{\left\{4\sqrt{2}\,\pi\,B(3/2,n+1/2) \right\}^{(n-3/2)/(n-1/2)}}\,\,
\left\{\frac{(n+1)^{3/2}}{h^3}\right\}^{1/(n-1/2)}.
\end{equation}
Equations (\ref{eq: beta-P_relation}) 
and (\ref{eq: Zq-rho_relation}) play a crucial role 
in determining the thermodynamic temperature of stellar polytrope 
in section \ref{sec: thermodynamics}, 
as well as the onset of gravothermal instability in section 
\ref{sec: stability_variation}.  
In particular, the equation (\ref{eq: beta-P_relation}) 
yields the {\it radius-mass-temperature relation} characterizing the 
quasi-equilibrium structure of the system confined in a thermal bath.

Keeping the above remarks in mind, hereafter, we will specifically 
focus on the spherically symmetric case with the 
polytrope index $n>3/2\,(q>0)$, in which the quasi-equilibrium distribution 
is at least dynamically stable (see Chap.5 of Ref.\cite{BT1987}). 
In this case, despite the above detailed 
differences,  the stellar quasi-equilibrium distribution 
can be characterized by the so-called {\it Emden solutions} 
(e.g., \cite{Chandra1939,KW1990}) and all the physical 
quantities are expressed in terms of the homology invariant 
variables $(u,v)$, which are subsequently used in later analysis. 
In Appendix B, together with some 
useful integral formulae, we summarize the relation between the Emden 
solution and the stellar polytropic distribution.
%
%
%
%
%
%
%
%
%
%
%
%
%
%
%
%
\section{Thermodynamic properties of stellar polytrope}
\label{sec: thermodynamics}
%
%
%
%
%
%
%
%
Having established the quasi-equilibrium distribution, we now 
investigate the thermodynamic properties of stellar polytropic system. 
To do this, we adopt the same procedure as in paper II.   
That is, we examine the Clausius relation under the quasi-static variation 
for the new equilibrium system in section \ref{subsec: clausius_relation}. 
In contrast to the previous analysis, 
the thermodynamic temperature can be identified with a help of 
(\ref{eq: beta-P_relation}), consistently with the recent claim based on 
the thermodynamic zero-th law. Using this temperature, in section 
\ref{subsec: negative_specific_heat}, presence or absence of thermodynamic 
instability is discussed evaluating the specific heat of stellar system. 
%
%
%
%
%
%
%
\subsection{On the definition of thermodynamic temperature}
\label{subsec: clausius_relation}
%
%
%
%
%
%
%
%
%
%
%
%
As usual, the thermodynamic instability in stellar system is 
intimately related to the presence of negative specific heat 
\cite{LW1968,LyndenBell1999}. 
The evaluation of specific heat is thus necessary for clarifying the 
thermodynamic property. In this regard, the identification of 
temperature in stellar system is the most essential task.

In the new framework of Tsallis' non-extensive thermostatistics, 
the physically plausible thermodynamic temperature, $\Tphys$, can be 
defined from the zero-th law of thermodynamics 
\cite{AMPP2001,Abe2001,MPP2001}. Then, the thermodynamic 
temperature in non-extensive system differs from 
the usual one, i.e., the inverse of the Lagrange multiplier, $\beta$. 
Depending on the methods extremizing the entropy, one has 
\begin{eqnarray}
\Tphys=\left\{
\begin{array}{cr}
\beta^{-1}                      & ;\,\,\,\,\,\,(\mbox{OLM method}), 
\\
\,[1+(1-q)\,S_q]\,\,\beta^{-1}  & ;\,\,\,\,\,\,(\mbox{TMP method}). 
\end{array}
\right.
\label{eq: def_Tphys}
\end{eqnarray}

Translating the above result into our notation immediately yields 
that the physical temperature in stellar polytropic system is given by 
$\Tphys=\Zq'/\beta$. However, the relation (\ref{eq: def_Tphys}) 
should be carefully applied to the present case, since the  
verification of thermodynamic zero-th law is very difficult 
in stellar equilibrium system with long-range interaction. 
Furthermore, even using the new formalism, the energy $E$ 
still keeps non-extensive due to the self-referential form of the 
potential energy (see Eqs.(\ref{eq: def_E})(\ref{eq: def_Phi})). 
In order to validate the use of the definition (\ref{eq: def_Tphys}), 
we therefore adopt a rather simple procedure as examined in paper II. 
That is, we consider the relation between heat transfer and entropy change 
in the quasi-static treatment under keeping the total mass constant. 
According to the definition (\ref{eq: def_Tphys}), the Clausius relation 
is appropriately modified and is expressed as follows 
\cite{AMPP2001,Abe2001,MPP2001}: 
\begin{equation}
  \label{eq: modified_clausius}
  dS_q = \frac{1}{\Tphys}\{ 1+(1-q)\,S_q\}\,d'Q. 
\end{equation}
This, in turn, implies that the modified Clausius relation 
can be used as a consistency check of the physical 
temperature (\ref{eq: def_Tphys}) in stellar system.

Let us first write down the entropy of extremum state:  
\begin{equation} 
S_q = \left(n-\frac{1}{2}\right)\,\,(\Zq-1).
\label{eq: S_q_extreme}
\end{equation} 
>From equation (\ref{eq: Zq-rho_relation}) with the polytropic relation 
(\ref{eq: polytrope}), the quantity $\Zq$ in equation 
(\ref{eq: S_q_extreme}) is rewritten as 
\begin{eqnarray}
  \Zq &=& c_n \, K_n^{n/(n-1/2)}\,
  \left\{\,\int d^3x\,\,P(x)\,\right\}^{-(n-3/2)(n-1/2)}
\nonumber \\
 &=& c_n \, K_n^{n/(n-1/2)}\, \left\{ \,\frac{1}{n-5}\,\frac{GM^2}{r_e}\,
 \left(\frac{n+1}{\ve}-2\frac{\ue}{\ve}-1\right)\right\}^{-(n-3/2)/(n-1/2)}
\end{eqnarray}
with a help of the integral formula (\ref{eq: formula_1}). 
Here, the variables $(\ue,\ve)$ 
denote the homology invariants evaluated at the boundary $r=\re$ 
(see definitions (\ref{eq: def_u})(\ref{eq: def_v})).  
Note also the fact that the constant $K_n$ is 
expressed in terms of $(\ue,\ve)$-variables: 
\begin{eqnarray}
  K_n = \frac{P_e}{\rho_e^{1+1/n}} = 
   \frac{GM}{\re}\,\frac{1}{\ve}\,\left(\frac{4\pi\re^3}{M}\,
  \frac{1}{\ue}\right)^{1/n}.
\end{eqnarray}
Using these expressions, the variation of the quantities 
$\Zq$ and $K_n$ respectively becomes 
\begin{eqnarray}
&&\left(n-\frac{1}{2}\right)\,\frac{d\Zq}{\Zq}\, =
\,n\,\frac{dK_n}{K_n}\,+\,\left(n-\frac{3}{2}\right)\,\frac{d\re}{\re}
\nonumber \\
&&~~~~~~~~~~~~  -\frac{n-3/2}{n-5}\,\frac{GM^2\beta}{\re\Zq'}\,
\left\{-\frac{n+1}{\ve}\,
  \frac{d\ve}{\ve}-2\left(\frac{d\ue}{\ue}-\frac{d\ve}{\ve}\right)
\right\},
\label{eq: dZ_q}
\end{eqnarray}
and 
\begin{eqnarray}
&&\frac{dK_n}{K_n} = -\frac{n-3}{n}\,\frac{d\re}{\re}\,-\frac{d\ve}{\ve}
  -\frac{1}{n}\,\frac{d\ue}{\ue}.
\label{eq: dK_n}
\end{eqnarray}
In the last line of equation (\ref{eq: dZ_q}),   
we have used the relation (\ref{eq: beta-P_relation}).  
Collecting these results, 
the entropy change $dS_q$ can be expressed as the variations of both
the homology invariants $(\ue,\ve)$ and the wall radius $\re$ 
as follows:
\begin{eqnarray}
&&  dS_q = \left(n-\frac{1}{2}\right)\,d\Zq 
  = \Zq\,\,\left[ \frac{3}{2}\,\frac{d\re}{\re}\,-
    \,\frac{d\ue}{\ue}\,-\,n\,\frac{d\ve}{\ve} 
\right.
\nonumber\\
&&~~~~~~~~~~~~~~\left. -\frac{n-3/2}{n-5}\,\frac{GM^2\beta}{\re\Zq'}\,
\left\{-\frac{n+1}{\ve}\,
  \frac{d\ve}{\ve}-2\left(\frac{d\ue}{\ue}-\frac{d\ve}{\ve}\right)
\right\}
\right], 
\end{eqnarray}
In the above equation, the term $GM^2\beta/(\re\Zq')$ can be 
factorized out using the relation (\ref{eq: beta-P_relation}) and 
we have 
\begin{eqnarray}
&&  dS_q = \Zq\,\,\frac{GM^2\beta}{\re\,\Zq'}
\,\,\left[\,-\,\frac{3}{2}\,\frac{1}{n-5}\,
\left( 2\,\frac{\ue}{\ve}-\frac{n+1}{\ve} + 1\right)\frac{d\re}{\re}
\right.
\nonumber\\
&&~~~~~~~~~~~~
+\frac{1}{n-5}\,\frac{1}{2\ve}\,
\left\{4\left(n-\frac{1}{2}\right)\ue+2\ve-2(n+1)\right\}\frac{d\ue}{\ue}
\nonumber\\
&&\left.~~~~~~~~~~~~~~~~~~~~~~~~~~~~~~~~~~~~~~~~
+\left\{6\ue+2n\ve-3(n+1)\right\}\frac{d\ve}{\ve}
\right]. 
\end{eqnarray}
Further notice the fact that the variation of homology invariants is 
expressed as the variation of dimensionless quantity $d\xxie$ 
(see Eq.(\ref{eq: d(u,v)/dxi})).  
The entropy change is finally reduced to the following expression: 
\begin{eqnarray}
  &&  dS_q = \Zq\,\,\frac{GM^2\beta}{\re\,\Zq'}
\,\,\left[\,-\,\frac{3}{2}\,\frac{1}{n-5}\,
\left( 2\frac{\ue}{\ve}-\frac{n+1}{\ve} + 1\right)\frac{d\re}{\re}
-\frac{n-2}{n-5}\,\frac{1}{2\ve}\right.
\nonumber\\
&&\left.\times\left\{
4\ue^2 + 2\ue\ve-\left(8+3\frac{n+1}{n-2}\right)\ue -
\frac{3}{n-2}\ve+3\left(\frac{n+1}{n-2}\right)\right\}
\frac{d\xi_e}{\xi_e}\,\right]. 
\label{eq: entropy_change} 
\end{eqnarray}

The Clausius relation in a quasi-static variation relates 
the variation of entropy with the heat change. According to the 
first-law of thermodynamics, the heat change in a quasi-static variation 
is estimated as follows (see Eq.(26) in paper II): 
\begin{eqnarray}
  d'Q \,\,= \,\,dE+ P_e dV 
&=& d\left(-\lambda\,\frac{
GM^2}{r_e}\right)+ 4\pi\re^2\,P_e\, d\re
\nonumber \\
  &=& \frac{GM^2}{\re}\,\,
\left\{\,\left(\lambda+\frac{\ue}{\ve}\right)\frac{d\re}{\re}-
\xxie\,\frac{d\lambda}{d\xxie}\frac{d\xxie}{\xxie}\right\},
\label{eq: heat_change}
\end{eqnarray}
where the dimensionless quantity $\lambda$ related to the energy 
and its derivative $d\lambda/d\xxie$ are respectively given by 
(see Eq.(\ref{eq: energy_uv})): 
\begin{eqnarray}
&&  \lambda\equiv -\frac{\re\,E}{GM^2}\,=
-\frac{1}{n-5}\,\left\{ \frac{3}{2}\,\left(1-\frac{n+1}{\ve}\right)
+(n-2)\frac{\ue}{\ve}\right\},
\label{eq: def_lambda}
\end{eqnarray}
and 
\begin{eqnarray}
&&\,\,\xxie\frac{d\lambda}{d\xxie} = \frac{n-2}{n-5}\,
\frac{g(\ue,\ve)}{2\ve}\,\,;\,\,
\nonumber\\
&& g(u,v) = 4u^2 +2uv-\left\{8+3\left(\frac{n+1}{n-2}\right)
\right\}u-\frac{3}{n-2}\,v+3\left(\frac{n+1}{n-2}\right).
\label{eq: g(u,v)}
\end{eqnarray}

Hence, the comparison between (\ref{eq: entropy_change}) and 
(\ref{eq: heat_change}) with (\ref{eq: def_lambda})(\ref{eq: g(u,v)}) 
leads to the modified Clausius relation corresponding to 
the equation (\ref{eq: modified_clausius}): 
\begin{eqnarray}
  dS_q = \frac{\beta}{\Zq'}\,\Zq\,(dE + P_e\,dV)
= \frac{\beta}{\Zq'}\,\,\left\{1+(1-q)\,S_q\right\}d'Q.   
\end{eqnarray}
%
%
%
%
%
%
%
%
Therefore, with this relation, the thermodynamic temperature can be 
consistently identified along the line of the argument in 
Ref.\cite{AMPP2001} and the plausible physical temperature is now 
\begin{equation}
  \label{eq: T_phys}
  \Tphys= \frac{\Zq'}{\beta}. 
\end{equation}
%
%
%
%
%
%
%
%
%
%
%
%
%
%
%
%
%
%
\subsection{Thermodynamic instability arising from 
the negative specific heat} 
\label{subsec: negative_specific_heat}
%
%
%
%
%
%
%
%
%
%
%
%
%
Once adopting the definition (\ref{eq: T_phys}), we immediately 
obtain the {\it radius-mass-temperature relation} in terms of the 
homology invariants as follows. Equation (\ref{eq: beta-P_relation}) 
in section \ref{sec: polytrope} leads to 
\begin{eqnarray}
\Tphys = \int d^3x\,P(x) = -\frac{1}{n-5}\,\frac{GM^2}{\re}\,
        \left(2\,\frac{\ue}{\ve}-\frac{n+1}{\ve}+1\right)
\label{eq: Tphys_P}
\end{eqnarray}
from (\ref{eq: formula_1}). 
Thus, defining the dimensionless quantity $\eta=GM^2/(\re\Tphys)$, we obtain 
  \begin{equation}
\label{eq: eta}
\eta\equiv\frac{GM^2}{r_e\Tphys} = \frac{(n-5)\,\ve}{n+1-2\ue-\ve}.    
  \end{equation}
This is in marked contrast to the result using the standard linear 
mean (c.f. Eq.(31) of paper II).  While the radius-mass-temperature relation 
in previous paper includes the residual dimensional parameter $h=(l_0v_0)^3$, 
the expression (\ref{eq: eta}) has no such parameter dependence and 
is quite similar to the result in Boltzmann-Gibbs case (e.g.,  Eq.(29) 
in Ref.\cite{LW1968} or Eq.(25) in Ref.\cite{Chavanis2002a})). 
The nice form of the radius-mass-temperature relation implies that the 
specific heat can be determined independently of the residual parameter 
$h$, which is indeed a desirable property for a rigid theoretical 
prediction without any uncertainty.

By definition, the specific heat at constant volume $\Cv$ is given by  
\begin{equation}
  \label{eq: specific_heat}
  \Cv \equiv \left(\frac{d E}{d\Tphys}\right)_e
  =\frac{\displaystyle \left(\frac{d E}{d\xi}\right)_e}
  {\displaystyle \left(\frac{d\Tphys}{d\xi}\right)_e}. 
\end{equation}
The numerator and the denominator in the last expression are 
respectively rewritten as follows:
\begin{eqnarray}
  \label{eq: stability_1}
& \left(\frac{d E }{d\xi}\right)_e =&-\frac{GM^2}{r_e}\frac{d\lambda}{d\xie}
=-\frac{GM^2}{r_e}\,\frac{n-2}{n-5}\,\,\frac{g(\ue,\ve)}{2\ue\xie},
\end{eqnarray}
and
\begin{eqnarray}
&\left(\frac{d\Tphys}{d\xi}\right)_e=& 
\frac{GM^2}{r_e}\frac{d}{d\xie}\,\eta^{-1}
=\frac{GM^2}{r_e}\,\frac{1}{n-5}\,\,\frac{k(\ue,\ve)}{\ue\xie}.
  \label{eq: stability_2}
\end{eqnarray}
In the above equations, the functions $g(\ue,\ve)$ is already 
given by (\ref{eq: g(u,v)})  
and the function $k(\ue,\ve)$ is expressed as 
\begin{eqnarray}
  \label{eq: h(u,v)}
&&  k(\ue,\ve) = 4\ue^2 +2\ue\ve - (n+9)\ue-\ve + n+1,   
\end{eqnarray}
from (\ref{eq: d(u,v)/dxi}). Thus, the expression of specific heat 
can be reduced to the following simple form: 
\begin{equation}
  \label{eq: C_v}
  \Cv = -\frac{n-2}{2}\,\,\,\,\frac{g(\ue,\ve)}{k(\ue,\ve)}. 
\end{equation}
Provided the homology invariants at the boundary, 
the specific heat in the new formalism 
is uniquely determined, free from the residual parameter $h$ 
(c.f. Eq.(42) in paper II). 
In this sense, the result (\ref{eq: C_v}) can be regarded 
as a successful outcome of the new Tsallis formalism using 
the normalized $q$-values.

Now, we focus on the thermodynamic instability inferred from 
the qualitative behavior of specific heat. Recall that the 
thermodynamic instability appears when the specific heat of the 
system changes its sign. From (\ref{eq: C_v}), we readily expect 
the two possibilities. One is the case when the function $g(\ue,\ve)$ 
changes its sign. In this case, the condition for marginal stability, 
$\Cv=0$, becomes 
\begin{equation}
  \label{eq: marginal_stability(1)}
  g(\ue,\ve)=0.
\end{equation}
The other cases appear when the sign of the function $k(\ue,\ve)$ 
is changed. In this case, the marginal stability leads to 
the divergent behavior, $\Cv\to\pm\infty$ and the condition is given by  
\begin{equation}
  \label{eq: marginal_stability(2)}
  k(\ue,\ve)=0.
\end{equation}

\begin{figure}
  \begin{center}
    \includegraphics*[width=14cm]{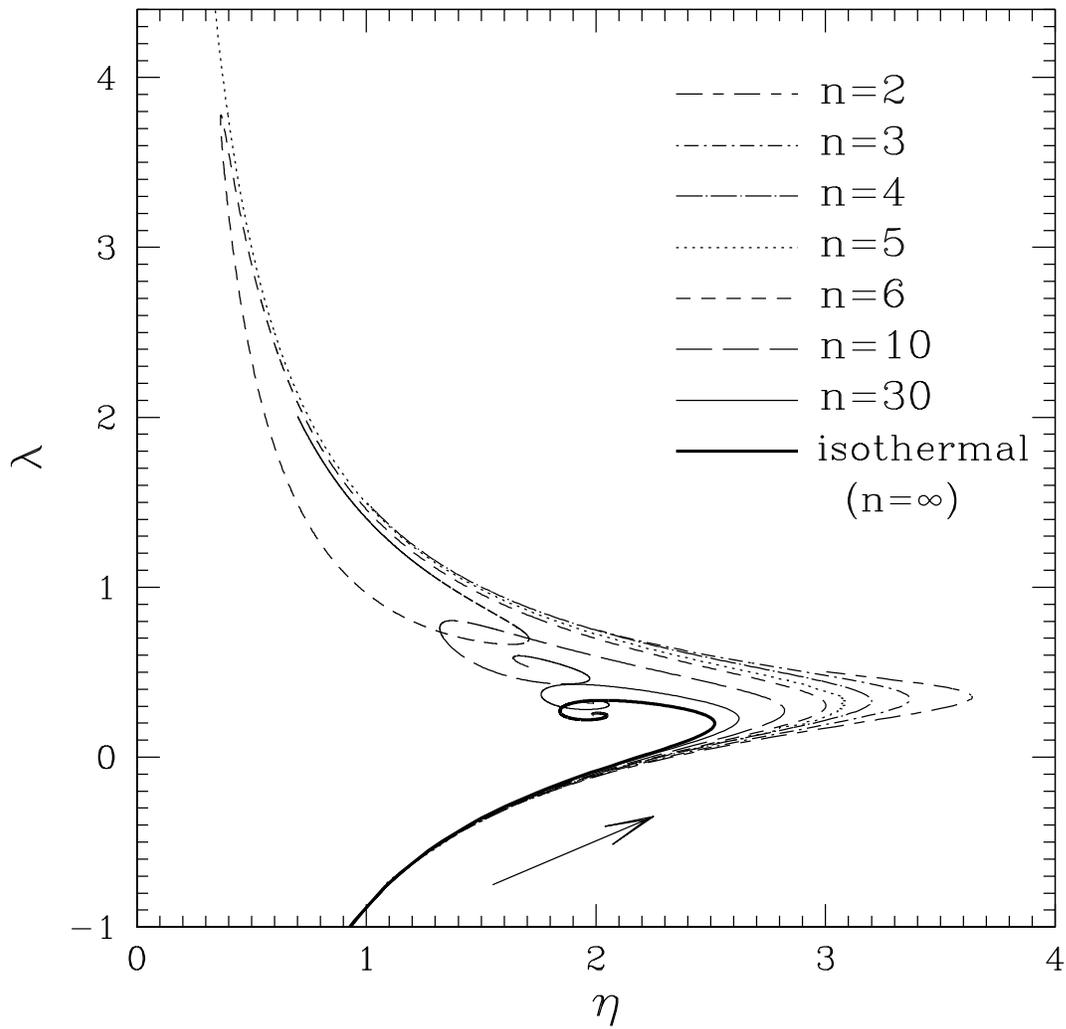}
    \caption{Trajectory of Emden solutions in $(\eta,\lambda)$-plane. }
    \label{fig: eta_lambda}
  \end{center}
\end{figure}

To see how and when the instability develops, 
in Fig. \ref{fig: eta_lambda}, we plot a family of Emden solutions in 
the $(\eta,\lambda)$-plane. Since the dimensionless parameters 
$\lambda$ and $\eta$ are respectively proportional to $-E$ and 
$\Tphys^{-1}$, the signature of the specific heat can be easily deduced 
from the slope of the curve. Note that each point along the trajectory 
represents an Emden solution evaluated at the different value 
of the radius $\re$. 
>From the boundary condition (\ref{eq: boundary}), 
all the trajectories start from $(\eta,\lambda)=(0,-\infty)$, corresponding 
to the limit $\re\to0$. As increasing the radius, the trajectories 
first move to the upper-right direction as marked by the arrow, and they 
suddenly change their direction to upper-left. This means that the divergent 
behavior of specific heat eventually appears and beyond that point, 
the signature of specific heat changes from positive to negative. 
That is, the potential energy dominates the kinetic energy($\lambda>0$) 
and the quasi-equilibrium state ceases to exist for a system in contact with a 
thermal bath (paper II). 
Notice that even in this case, the stable quasi-equilibrium state  
still exists for a system surrounded by an 
adiabatic wall. On the other hand, for more larger radius, 
while the curves with index $n<5$ abruptly terminate, 
the trajectories with $n>5$ next reach at another 
critical point $d\eta/d\lambda=0$, i.e., $\Cv=0$. Further, 
they progressively change their direction and finally spiral 
around a fixed point. 
The appearance of the critical point $C_{\rm V} = 0$ is explained as follows. 
While the inner part of the system keeps the 
specific heat negative, the outer part seems to have positive 
one. Thus, the heat current from inner to outer part causes 
the raise of the temperature at both parts. For a system with the radius 
$r_e$ smaller than certain critical value, the amount of the heat capacity 
at outer part is  small so that the outer part easily catches up with the 
increase of the inner-part temperature. 
As increasing $r_e$, the fraction of the outer normal part grows up 
and it eventually balances with the inner gravothermal part. 
Thus, beyond the point characterized by the condition $C_{\rm V} = 0$, 
no thermal balance is attainable and 
the system becomes gravothermally unstable. This is even true in the 
system surrounded by an adiabatic wall.

>From these discussions, one can immediately verify that the 
condition (\ref{eq: marginal_stability(1)})  represents the 
marginal stability for a system confined in an adiabatic 
wall, which exactly coincides with the previous result using standard 
linear means. Hence, one concludes that the quasi-equilibrium structure obtained 
from the new formalism does not alter the thermal properties in 
micro-canonical ensemble case. On the other hand, the condition 
(\ref{eq: marginal_stability(1)}) indicates the onset of thermodynamic 
instability for a system in contact with a thermal bath (i.e., canonical 
ensemble case), 
which significantly differs from the results in old Tsallis formalism. 
Of course, 
this might be a natural consequence of the different choice of 
statistical average, leading to the different definition $\Tphys$, 
however, the appearance of instability in present case 
might seem somewhat curious.  
While the previous results indicate 
the unstable state at the indices 
$n>3$, consistent with the suggestion by Chavanis 
\cite{Chavanis2002b}, 
Fig. \ref{fig: eta_lambda} implies that 
thermodynamic instability appears for stellar polytrope with 
any value of the index $n$. 
One might worry about whether the 
present results rigorously match the stability analysis from the 
variational problem. In next section, 
to check the consistency of new Tsallis formalism, we develop the 
stability analysis  based on the second variation of entropy and 
free-energy.

\begin{figure}
  \begin{center}
  \includegraphics*[width=14cm]{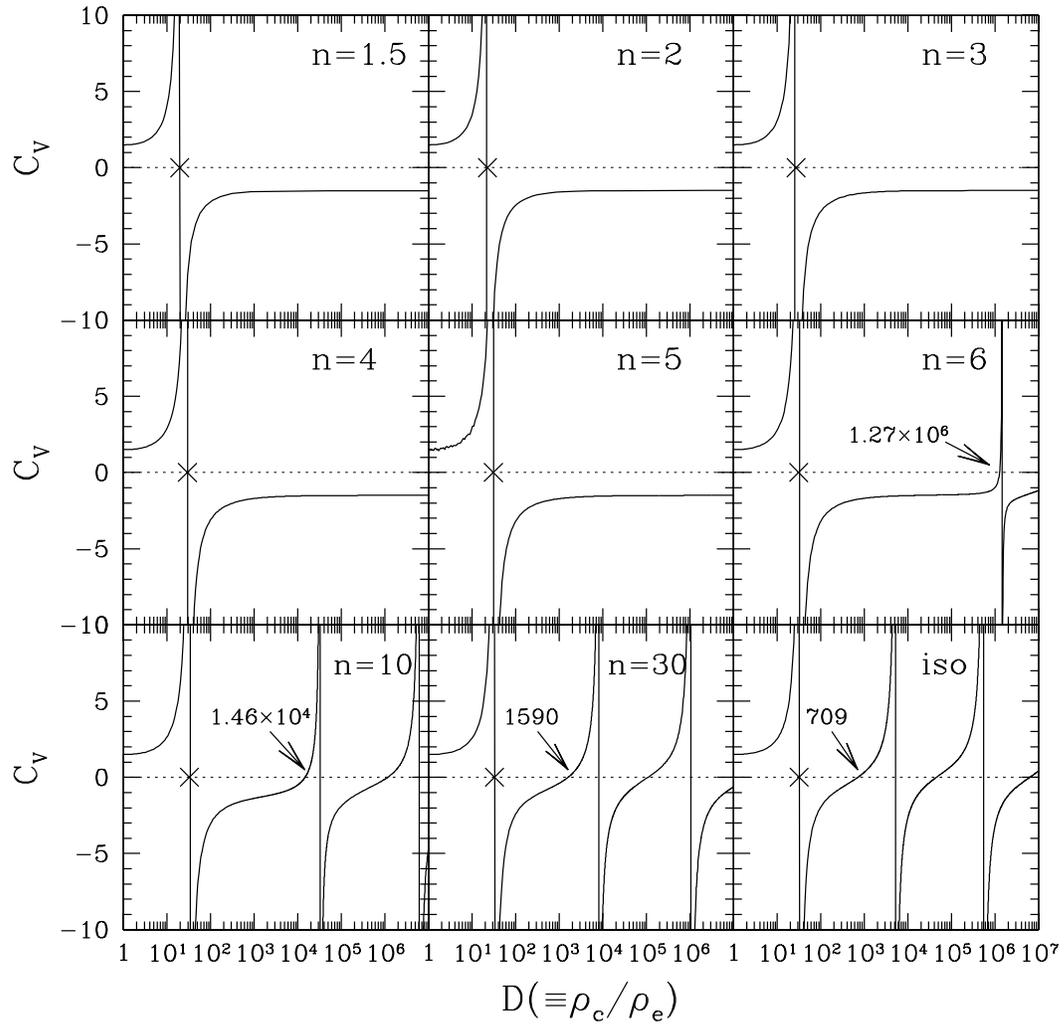}
  \end{center}
    \caption{Specific heat as a function of density contrast 
        $D(=\rho_c/\rho_e)$ for various polytrope indices. }
    \label{fig: c_v}
\end{figure}

\begin{figure}
  \begin{center}
    \includegraphics*[width=10.5cm]{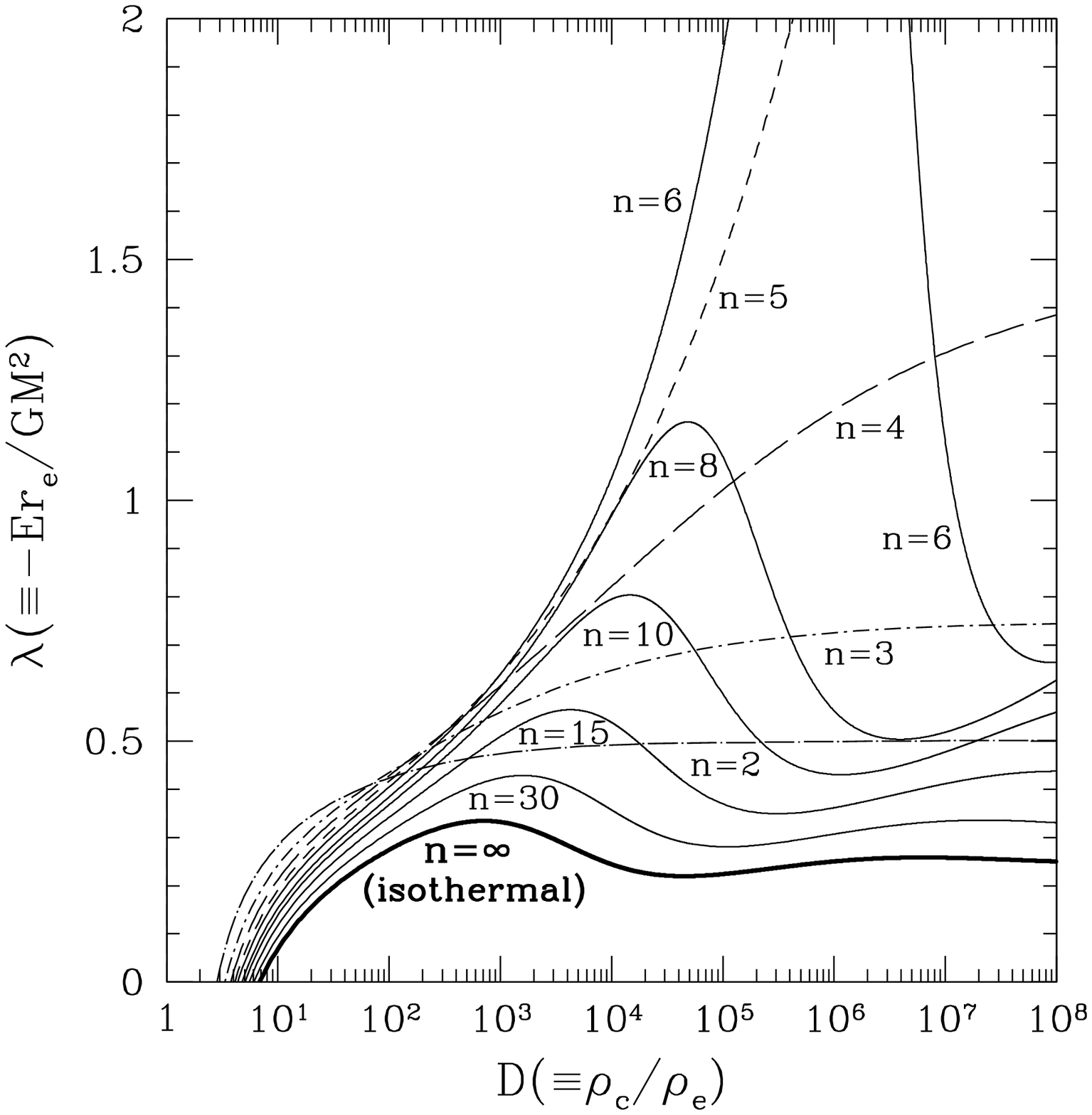}
    \caption{Energy-radius-mass relation as a function of density 
      contrast.}
    \label{fig: lambda_d}


    \includegraphics*[width=10.5cm]{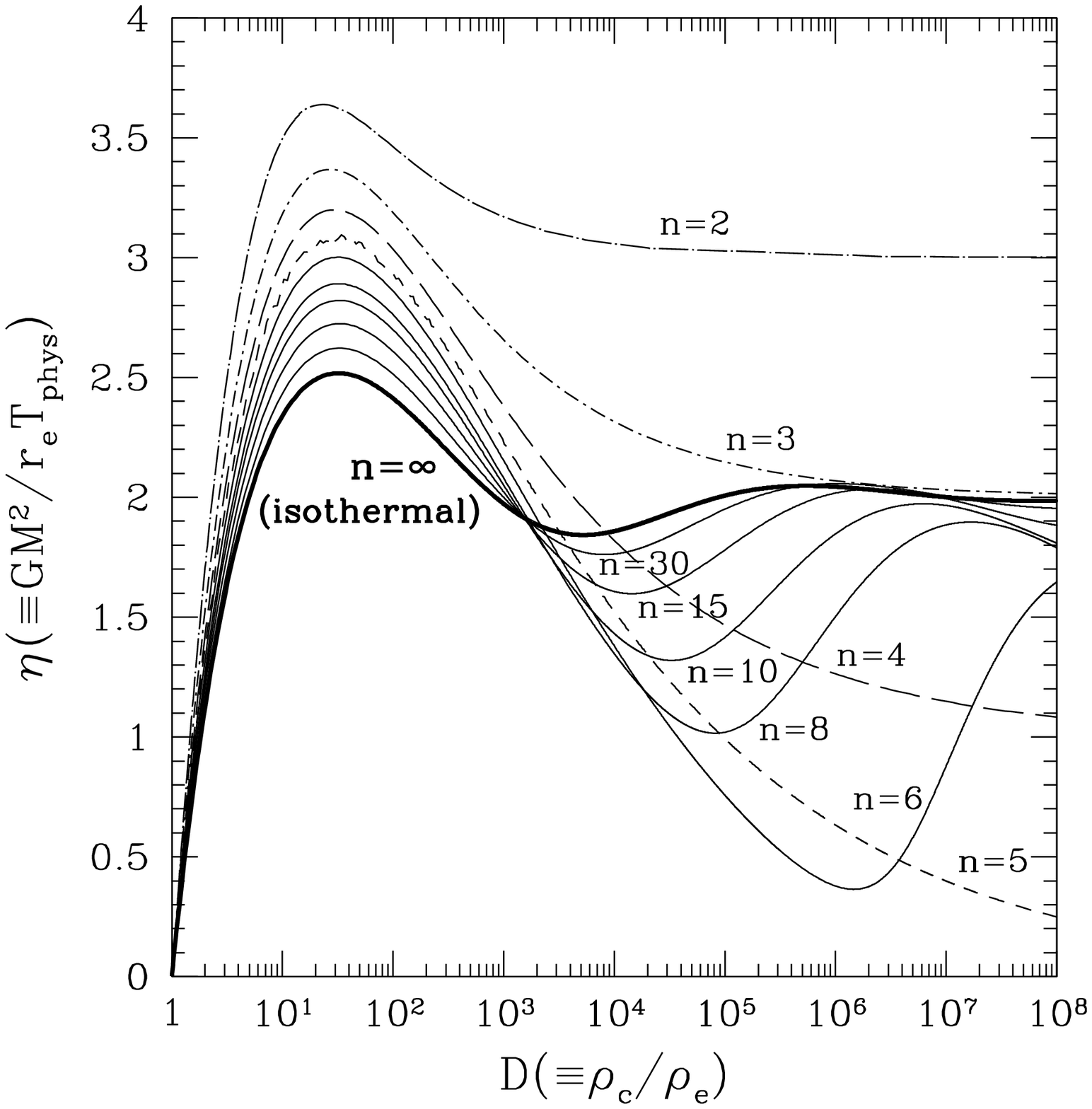}
    \caption{Radius-mass-temperature relation as a function of 
      density contrast.}
    \label{fig: eta_d}
  \end{center}
\end{figure}

Finally, in Fig. \ref{fig: c_v}, varying the radius $\re$, 
the specific heat $\Cv$ is plotted as a function of density 
contrast, $D\equiv\rho_c/\rho_e$ for typical polytrope indices 
with $n\geq 3/2$. Clearly, the critical point $|\Cv|\to\infty$ 
marked by crosses exists in each case, while the marginal stability 
$\Cv=0$ only appears when $n>5$ at a 
certain high density contrast (c.f. Fig. 3 in paper II). 
The numerical values indicated by arrows represent the 
critical values $D_{\rm crit}$ evaluated at the point $\Cv=0$, which are 
the same results as in Table 1 of paper I (see also Fig. 
\ref{fig: lambda_crit}). 
These behaviors can also be deduced from the energy-radius-mass relation 
and the radius-mass-temperature relation. In Figs. \ref{fig: lambda_d} 
and \ref{fig: eta_d}, using the expressions (\ref{eq: eta}) and 
(\ref{eq: def_lambda}), the dimensionless values $\lambda$ and $\eta$ 
for various polytrope indices are evaluated and plotted as a function of 
density contrast, respectively.\footnote{
Fig. \ref{fig: lambda_d} is essentially the same result as 
in Fig. 2 of paper I.} 
The Emden trajectories in $(\eta,\lambda)$ plane 
(see Fig.\ref{fig: eta_lambda}) state that the marginal stability for 
the system 
in contact with a thermal bath, $|\Cv|\to\infty$ implies  
the first turning point $d\eta/d\xxie=0$, or equivalently $d\eta/dD=0$. 
Similarly, the marginal stability $\Cv=0$ represents the condition 
$d\lambda/dD=0$. 
>From Figs. \ref{fig: lambda_d} and \ref{fig: eta_d}, we readily estimate 
the critical density contrast at the first turning points in each case, 
which exactly coincide with the points marked by the arrows and the 
crosses, respectively. Consistently, the turning point disappears 
when $n<5$ in Fig. \ref{fig: lambda_d}, while it does always exist 
independently of the polytrope index in Fig. \ref{fig: eta_d}. 
%
%
%
%
%
%
%
%
%
%
%
%
%
%
%
\section{Stability/instability criteria from the variational problems} 
\label{sec: stability_variation}
%
%
%
%
%
%
%
%
%
%
%
%
%
%
%
Previous section reveals the existence of two types of thermodynamic 
instability.  
Then, the marginal stability conditions (\ref{eq: marginal_stability(1)}) 
and (\ref{eq: marginal_stability(2)}) are obtained for 
a system confined in an adiabatic wall and for a system in contact with 
a thermal bath, respectively. 
In this section, in order to check the consistency of these results, 
we reconsider the marginal stability criteria based on the variational 
problem.

According to the maximum entropy principle, 
the stable quasi-equilibrium state for a system confined in an adiabatic wall 
is only possible when the second variation of entropy around the 
extremum state is negative, i.e.,  $\delta^2S_q<0$.  Similarly, the 
stable quasi-equilibrium distribution surrounded by a thermal bath should 
have minimum free-energy, indicating the positive value of the 
second variation of free-energy, i.e., $\delta^2F_q>0$. Thus, 
the condition $\delta^2S_q=0$ or $\delta^2F_q=0$ readily implies 
the marginal stability in each case.

To begin with, let us write down the entropy of quasi-equilibrium state. 
>From (\ref{eq: Zq-rho_relation}) and (\ref{eq: S_q_extreme}), 
we have 
\begin{equation}
  S_q = \left(n-\frac{1}{2}\right)\,\left\{
    c_n\,\,\left(\Tphys^{-(3/2)/(n-3/2)}\,\,W\right)^{-(n-3/2)/(n-1/2)}-1
\,\,\right\}.  
\label{eq: S_q_max}
\end{equation}
Here, just for convenience, we introduced the quantity $W$: 
\begin{equation}
  \label{eq: def_of_W}
  W\,=\,\int\,d^3 x \,\,\rho^{1+1/n}. 
\end{equation}
Note that in deriving the equation (\ref{eq: S_q_max}), we used the 
relation $\Tphys=K_n\,W$ from (\ref{eq: beta-P_relation}).

Using the above expression (\ref{eq: S_q_max}), 
we compute the variation of entropy up to the second order terms. 
To be specific, we consider the density perturbation $\delta\rho(r)$ 
around the quasi-equilibrium configuration, under keeping the total mass $M$ 
and the radius $\re$ constant. Then the variation of entropy becomes  
\begin{eqnarray}
&  \delta S_q =& \left(n-\frac{1}{2}\right)\,\,\delta \Zq
\nonumber\\
&    =& \Zq\,\,\left[\frac{3}{2}\frac{\delta\Tphys}{\Tphys}-
    n\frac{\delta W}{W}-
\frac{3}{4}\,\frac{n-2}{n-1/2}\,\left(\frac{\delta\Tphys}{\Tphys}\right)^2
\right.\nonumber\\
&&~~~~~~~~~~~~~\left.
+\frac{n(n-1/4)}{n-1/2}\,\left(\frac{\delta W}{W}\right)^2
-\frac{3}{2}\frac{n}{n-1/2}\,\frac{\delta\Tphys}{\Tphys}\frac{\delta W}{W}
\,\right]
\label{eq: d2S_q}
\end{eqnarray}
As for the variation $\delta W$, we have
\begin{eqnarray}
  \delta W = \delta\left(\int d^3x\,\,\rho^{1+1/n}\right)
  =\frac{n+1}{n}\,\int d^3x \left\{\,\delta\rho+\frac{1}{2n}\,
\frac{(\delta\rho)^2}{\rho}\,\right\}\rho^{1/n}.
\label{eq: d2W}
\end{eqnarray}
Also, we write down the variation of energy: 
\begin{eqnarray}
&&  \delta E = \delta\,\left(\frac{3}{2}\,\int d^3 x \,\,P + 
\frac{1}{2}\,\int d^3 x\,\, \rho\,\Phi\right)
\nonumber \\
&&~~~~~~~~~~~~~~~~~~~~~=\frac{3}{2}\,\delta\Tphys\,+\,
\frac{1}{2}\,\int d^3x\,\,\left(2\,\Phi\,\delta \rho\,+\,
\delta\rho\,\delta\Phi\,\right), 
\label{eq: d_E}
\end{eqnarray}
where we used the fact that 
$\int d^3x\,\rho\delta\Phi = \int d^3x\,\Phi\delta\rho$.

Below, we separately analyze the variational problem in each case.   
%
%
%
%
%
%
%
%
%
%
\subsection{Stability/instability criterion from the second 
variation of entropy}
\label{subsec: entropy}
%
%
%
%
%
%
%
%
Let us first consider the stability/instability condition for a 
system confined in an adiabatic wall. In this case, 
the conservation of total energy $E$ is always guaranteed and   
we further put another constraint $\delta E=0$, which yields  
\begin{eqnarray}
\delta\Tphys = -\frac{1}{3}\,\,\int d^3x\, 
        (2\Phi\,\delta\rho \,+\,\delta\rho\,\delta\Phi)
\nonumber
\end{eqnarray}
from (\ref{eq: d_E}).
Substituting the above equation into (\ref{eq: d2S_q}), 
a straightforward calculation leads to 
the variation of entropy up to the second order, 
summarized as follows: 
\begin{eqnarray}
  \delta S_q = \delta^{(1)}S_q + \delta^{(2)}S_q\,\,\, ;
\nonumber 
\end{eqnarray}
\begin{eqnarray}
  \delta^{(1)} S_q \,&= &\,-\, \frac{\Zq}{\Tphys}\,\,
\int d^3x\left\{\, \Phi + 
(n+1)\,\frac{\Tphys}{W}\,\rho^{1/n}(x)\,\right\}\delta\rho,
\label{eq: del(1)S_q}   
\end{eqnarray}
\begin{eqnarray}
  \delta^{(2)} S_q \,&=&\,-\Zq\, \left[\,\,
\int d^3x\left\{ \frac{1}{2\,\Tphys}\,\delta\rho\,\delta\Phi + 
\frac{1}{2W}\,\frac{n+1}{n}\,\rho^{1/n-1}(\delta\rho)^2\,\right\}
\right.
\nonumber \\
&+&\frac{1}{3\Tphys^2}\,\frac{n-2}{n-1/2}\,
\left(\int d^3 x\,\Phi\,\delta\rho\right)^2 
\nonumber \\
&& ~~-\frac{1}{W^2}\,\frac{(n+1)^2(n-1/4)}{n(n-1/2)}\,
\left(\int d^3 x\,\delta\rho^{1/n}\,\delta\rho\,\right)^2 
\nonumber \\
&& \left.~~~~~-\frac{1}{W\,\Tphys}\,\frac{n+1}{n-1/2}\,
\left(\,\int d^3x\,\rho^{1/n}\delta\rho\,\right)
\left(\,\int d^3x\,\Phi\,\delta\rho\,\right)\,\,\right].
\label{eq: del(2)S_q}   
\end{eqnarray}

The resultant form of the second variation of entropy $\delta^{(2)} S_q$ 
seems rather complicated, however, recalling  the fact that 
the background solution of stellar polytropic distribution always satisfies 
the condition $\delta^{(1)}S_q=0$, the last three terms in right-hand-side 
of equation (\ref{eq: del(2)S_q}) can be rewritten in more compact form. 
After some algebra, we obtain
\begin{eqnarray}
&&  \delta^{(2)}S_q = -\,\Zq\,\left[ \,\,\int d^3x \,
  \left\{ \frac{1}{2\Tphys}\,\delta\rho\,\delta\Phi\,+\,
  \frac{1}{2W}\,\frac{n+1}{n}\,\rho^{1/n-1}\,(\delta\rho)^2\right\}
\right.
\nonumber \\
&&~~~~~~~~~~~~~
\left. +\frac{W}{3\Tphys^2}\,\,\frac{n}{n-3/2}\,
\left(\int d^3x\,\Phi\,+\,\frac{3}{2}\frac{n+1}{n}\,\frac{\Tphys}{W}\,\,
\delta\rho\right)^2\,\,\right]. 
\end{eqnarray}
Apart from the over-all positive constant, the above expression is 
indeed the same equation as previously obtained in paper I (see Eq.(36) 
in paper I, with the identification $\Tphys/W=\{(n-3/2)/(n+1)\}T$). 
Thus, just following the same calculation as in paper I, 
the stability/instability criterion from $\delta^{(2)}S_q=0$ can be 
obtained, from which one can rigorously prove that the marginal stability 
condition 
for a system confined within an adiabatic wall exactly reproduces the equation 
(\ref{eq: marginal_stability(1)}). As a result, for certain critical 
values of dimensionless energy and density, $\lambda_{\rm crit}$ and 
$D_{\rm crit}$, the onset of the gravothermal instability appears when 
the stellar polytrope with index $n>5$ (see Figs.\ref{fig: lambda_d} 
and \ref{fig: lambda_crit}).
%
%
%
%
%
%
%
%
%
%
\subsection{Stability/instability criterion from the second variation of 
free-energy}
\label{subsec: free-energy}
%
%
%
%
%
%
%
%
%
%
Now, turn to focus on the second variation of free-energy. 
      In paper II, usual definition of free-energy 
\begin{equation}
F'_q = E - \beta^{-1} S_q 
\label{eq: old_free-energy}
\end{equation}
has been used to investigate the marginal stability for a system surrounded 
by a thermal bath. As several authors recently pointed out, the usual 
definition 
of free-energy becomes inadequate when the physical temperature cannot be 
identified with the inverse of Lagrange multiplier $\beta$. 
According to the modification of Clausius 
relation (\ref{eq: modified_clausius}), the free-energy must be generalized 
as follows \cite{AMPP2001,Abe2001}: 
\begin{equation}
F_q =  E - \Tphys\,\frac{1}{1-q}\ln\left\{1+(1-q)S_q\right\}.
\label{eq: generalized_free-energy}
\end{equation}
Using this {\it generalized} free-energy,  we consider the 
variation up to the second order, keeping the thermodynamic temperature 
$\Tphys$ constant, not the total energy $E$:    
\begin{eqnarray}
\delta F_q = \delta E - \Tphys\,\frac{1}{1-q}\,
\left\{
\frac{\delta \Zq}{\Zq}-\frac{1}{2}\left(\frac{\delta \Zq}{\Zq}\right)^2
\right\}.
\nonumber
\end{eqnarray}
Substituting the equations (\ref{eq: d2S_q}), (\ref{eq: d2W}) 
and (\ref{eq: d_E}) 
into the above equation and keeping the terms up to the second order, 
after some manipulation, one obtains
\begin{eqnarray}
  \delta F_q = \delta^{(1)}F_q + \delta^{(2)}F_q\,\,\,;
\nonumber
\end{eqnarray}
\begin{eqnarray}
\delta^{(1)} F_q &=& \int d^3x\left\{\Phi(x) + 
(n+1)\frac{\Tphys}{W}\rho^{1/n}(x)\right\}\delta\rho,
\label{eq: del(1)F_q}   
\end{eqnarray}
\begin{eqnarray}
\delta^{(2)}F_q &=& 
\frac{1}{2} \int d^3x \left\{ \delta\rho \delta\Phi + 
\frac{n+1}{n}\, \frac{\Tphys}{W}\,\rho^{1/n-1}(\delta\rho)^2 \right\} 
\nonumber\\
&&
~~~~~~~~~~~~~~~~~~~~~~~
-\frac{1}{2}\,\frac{\Tphys}{n}\,\left(\frac{n+1}{W}\right)^2\,
\left(\int d^3x \rho^{1/n}\delta\rho\right)^2.
\label{eq: del(2)F_q}   
\end{eqnarray}
Note that apart from the over-all factor, the first 
variation $\delta^{(1)}F_q$ exactly coincides 
with the first variation of entropy, $\delta^{(1)}S_q$, which ensures the 
fact that extremum state of generalized free-energy is the same 
quasi-equilibrium distribution as obtained from the maximum entropy 
principle. Using the fact $\delta^{(1)}F_q=0$, equation (\ref{eq: del(1)F_q}) 
is now rewritten with 
\begin{eqnarray}
&&\delta^{(2)}F_q = 
\frac{1}{2} \int d^3x \left\{ \delta\rho \delta\Phi + 
\frac{n+1}{n} \,\frac{\Tphys}{W}\, \rho^{1/n-1}(\delta\rho)^2 \right\} 
\nonumber \\
&&~~~~~~~~~~~~~~~~~~~~~~~~~~~~~~~~~~~~~~~~~~~~~~~~
 -\frac{1}{n\Tphys}\,
\left(\int d^3x \Phi\delta\rho\right)^2.
\label{eq: reduced_del(2)F_q}
\end{eqnarray}
Obviously, the above equation differs from the old result based on the 
usual definition of free-energy (\ref{eq: old_free-energy}) (c.f. Eq.(51) 
in paper II). This fact readily implies that the thermodynamic stability 
has nothing to do with the dynamical stability, which apparently contradicts 
with the recent claim by Chavanis \cite{Chavanis2002a,Chavanis2002b}. 
Thus, in present case using normalized $q$-values, it seems non-trivial 
whether the above equation indeed leads to the marginal stability condition 
(\ref{eq: marginal_stability(2)}).

In what follows, restricting our attention to the radial 
mode of density and potential 
perturbations, we discuss the existence or the absence of perturbation 
mode satisfying the equation $\delta^{(2)}F_q=0$. 
Following the papers I and II, we introduce the new perturbed quantity: 
\begin{equation}
  \label{eq: Q}
  \delta\rho(r)\equiv \frac{1}{4\pi r^2}\frac{dQ(r)}{dr}. 
\end{equation}
Then the mass conservation $\delta M=0$ 
implies the boundary condition $Q(0)=Q(r_e)=0$. 
Substituting (\ref{eq: Q}) into (\ref{eq: reduced_del(2)F_q}) 
and repeating the integration by parts, 
the second variation of generalized free-energy 
is reduced to the following quadratic form: 
\begin{equation}
  \label{eq: del_2_F_}
  \delta^{(2)}F_q = -\frac{1}{2} \int_{0}^{r_e} dr_1\,\int_{0}^{r_e} dr_2\,\,
Q(r_1)\, \hat{K}(r_1, r_2) \, Q(r_2),
\end{equation}
where $\hat{K}(r_1, r_2)$ stands for the operator given by 
\begin{eqnarray}
  \label{eq: operator_K}
  \hat{K}(r_1, r_2) &\equiv& \left\{ \frac{n+1}{n}\,\frac{d}{dr_1}
\left(\frac{1}{4\pi r_1^2\rho(r_1)}\frac{P(r_1)}{\rho(r_1)}\frac{d}{dr_1}\right)+
\frac{G}{r_1^2}\right\}\delta_D(r_1-r_2)
\nonumber\\
&&~~~~~~~~~~~~~~~~~~~~~~~~~~~~~~~~~~
+\frac{1}{n\Tphys}\,\frac{d\Phi(r_1)}{dr_1}\frac{d\Phi(r_2)}{dr_2}.
\end{eqnarray}

Thus, the problem reduces to the eigenvalue problem and 
the marginal stability condition just corresponds to 
the zero-eigenvalue problem: 
\begin{eqnarray}
\int_0^{\re}dr' \,\hat{K}(r,r')\,\,Q(r') =  0, 
\end{eqnarray}
which yields
\begin{eqnarray}
  \label{eq: zero-eigenvalue_eq}
&&  \hat{L}[Q]\equiv 
\left[\frac{n+1}{n}\,\frac{d}{dr}\left(\frac{1}{4\pi\,r^2\,\rho}\,\,
\frac{P}{\rho}\,\frac{d}{dr}\right)+\frac{G}{r^2}\right]Q
\nonumber\\
  &&~~~~~~~~~~~~~~~~~~~~~~~~~~~~~~
=-\frac{1}{n\Tphys}\,\frac{G\,m(r)}{r^2}\,
        \int_0^{r_e}dr' \frac{Gm(r')}{r'^2}\,Q(r').
\end{eqnarray}

To obtain the solution of the above zero-eigenvalue problem, 
we follow the similar procedure in paper I. First recall the fact 
that the following equations are satisfied:  
\begin{equation}
\hat{L}[4\pi\,r^3\,\rho(r)]=\frac{n-3}{n}\,\frac{Gm(r)}{r^2},
~~~~~~~~~~
\hat{L}[m(r)]=\frac{n-1}{n}\,\frac{Gm(r)}{r^2}.
\end{equation}
Then, these relations allow us to put the ansatz of the solution, 
\begin{equation}
  Q(r) = c_1 \,4\pi r^3 \rho(r) + c_2 \, m(r), 
  \label{eq: ansatz}
\end{equation}
and to determine the coefficients $c_1$ and $c_2$ by 
substituting the ansatz into (\ref{eq: zero-eigenvalue_eq}):   
\begin{equation}
  \frac{n-3}{n}\,\,c_1 + \frac{n-1}{n}\,\, c_2  = \Lambda_n, 
        \label{eq: condition_1}
\end{equation}
where we define 
\begin{equation}
  \label{eq: Lambda_n}
  \Lambda_n \equiv -\frac{1}{n\Tphys}\,\int_0^{r_e}\,dr\,\,
  \frac{Gm(r)}{r^2}\,\,Q(r).
\end{equation}
In addition to the above condition, we must further consider 
the boundary conditions $Q(0)=Q(\re)=0$. 
While the condition $Q(0)=0$ is automatically fulfilled,  
the remaining condition $Q(r_e)=0$ requires 
\begin{equation}
\label{eq: condition_2}
  c_1 \,4\pi r^3 \rho_e + c_2 \, M = 0.
\end{equation}
Hence, from (\ref{eq: condition_1}) and (\ref{eq: condition_2}),  
the coefficients $c_1$ and $c_2$ are determined and are expressed 
in terms of homology invariants:  
\begin{equation}
  \label{eq: c_1,c_2}
  c_1 = \frac{n\,\,\Lambda_n}{n-3-(n-1)\,\ue},
~~~~~
  c_2 = -\frac{n\,\,\ue\,\Lambda_n}{n-3-(n-1)\,\ue}.
\end{equation}
As noted in paper I, the above coefficients are not specified completely 
because of the quantity $\Lambda_n$ in the coefficients, which depends on 
the solution (\ref{eq: ansatz}) itself.  To eliminate this self-referential 
structure, we directly evaluate the non-local term $\Lambda_n$, 
leading to the consistency condition for the existence of solution 
(\ref{eq: ansatz}). Substitution of the solution (\ref{eq: ansatz}) with 
the coefficients (\ref{eq: c_1,c_2}) into the expression (\ref{eq: Lambda_n}) 
yields
\begin{eqnarray}
  1= -\,\frac{\Tphys^{-1}}{n-3-(n-1)\ue}  \,\,\int_0^{r_e}\,dr\,\,
\frac{Gm(r)}{r^2}\,\,\left\{4\pi r^3\rho(r)-\ue\,m(r)\right\}.
\nonumber 
\end{eqnarray}
In the above equation, the integrals in the right-hand-side 
can be evaluated using 
the formulae (\ref{eq: formula_2}) and (\ref{eq: formula_3}) 
listed in Appendix B. Further, the radius-mass-temperature relation  
(\ref{eq: eta}) eliminates the dependence of $\Tphys$, leading to   
\begin{eqnarray}
&&  1 = \frac{\ve}{2\ue+\ve-(n+1)}\,\,\frac{1}{n-3-(n-1)\ue}
\nonumber\\
&&~~~~~~~~~~~~~\times \left[(n-5)\ue + (2\ue-1)
\left\{3+(n+1)\left(\frac{\ue}{\ve}-\frac{3}{\ve}\right)\right\}\right].
\nonumber
\end{eqnarray}
After some algebra, the above equation is rewritten with 
the following quadratic form: 
\begin{eqnarray}
  0&=&4\ue^2 + 2\ue\ve-(n+9)\ue-\ve+n+1 
\nonumber \\
&=& k(\ue,\ve),
\end{eqnarray}
which coincides with the marginally stability criterion 
(\ref{eq: marginal_stability(2)}) derived from 
the condition for specific heat, $\Cv\to\infty$.

Therefore, unlike the first impression, we reach at 
the fully satisfactory conclusion that the thermodynamic instability 
inferred from the specific heat is consistently explained from the 
variational problems. In other words, the application of 
new Tsallis formalism to the stability of stellar quasi-equilibrium 
system reveals a consistent thermodynamic structure of 
self-gravitating system, as well as the existence of thermodynamic 
instability. 
%
%
%
%
%
%
%
%
%
%
%
\section{Discussion \& conclusion}
\label{sec: conclusion}
\begin{figure}
  \begin{center}
\vspace*{-1.0cm}
\hspace*{0.6cm}
\vspace*{-1.0cm}
    \includegraphics*[width=12.5cm]{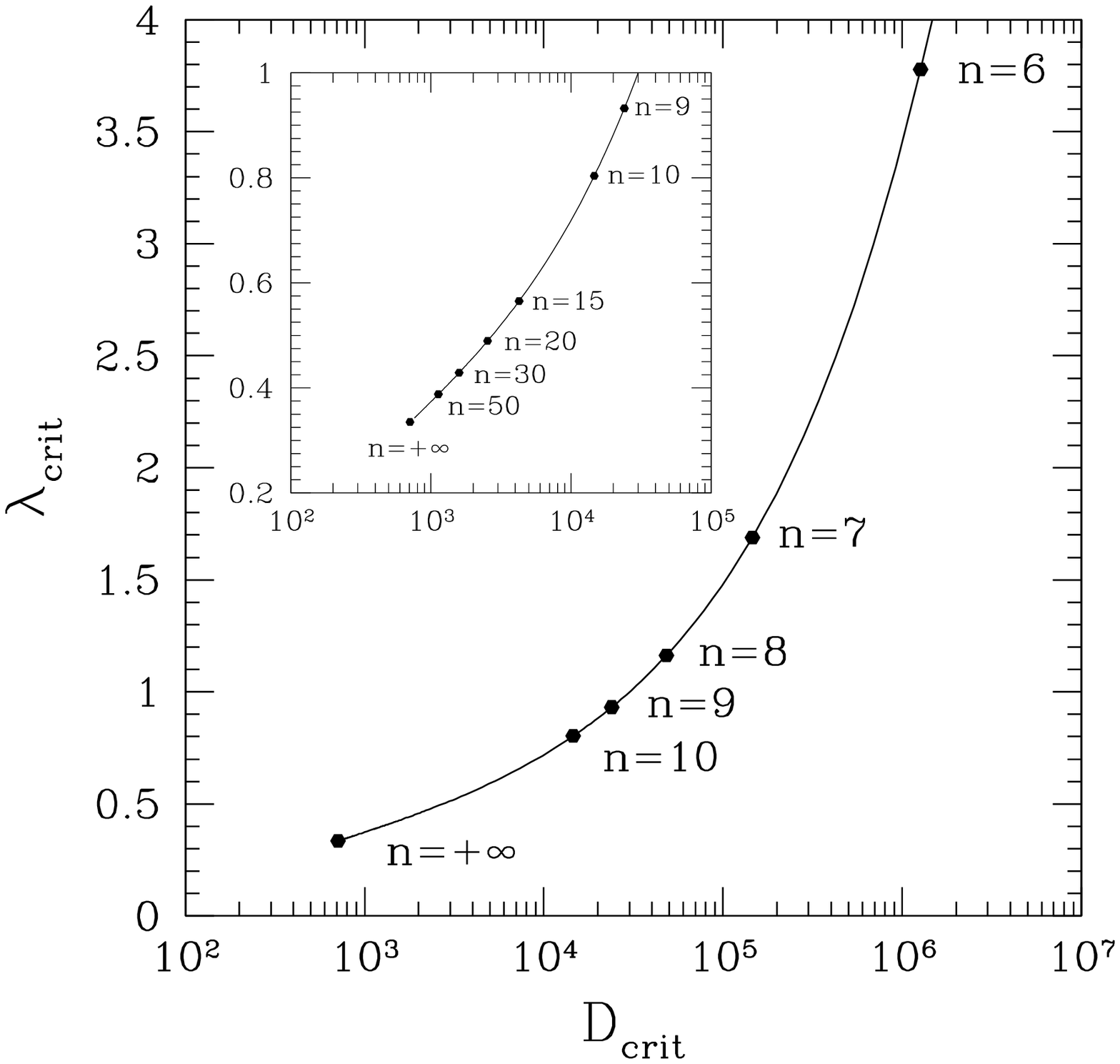}
    \caption{Critical values $(D_{\rm crit}, \lambda_{\rm crit})$ 
      for a system confined in an adiabatic wall.} 
    \label{fig: lambda_crit}

\vspace*{+0.8cm}

    \includegraphics*[width=10cm]{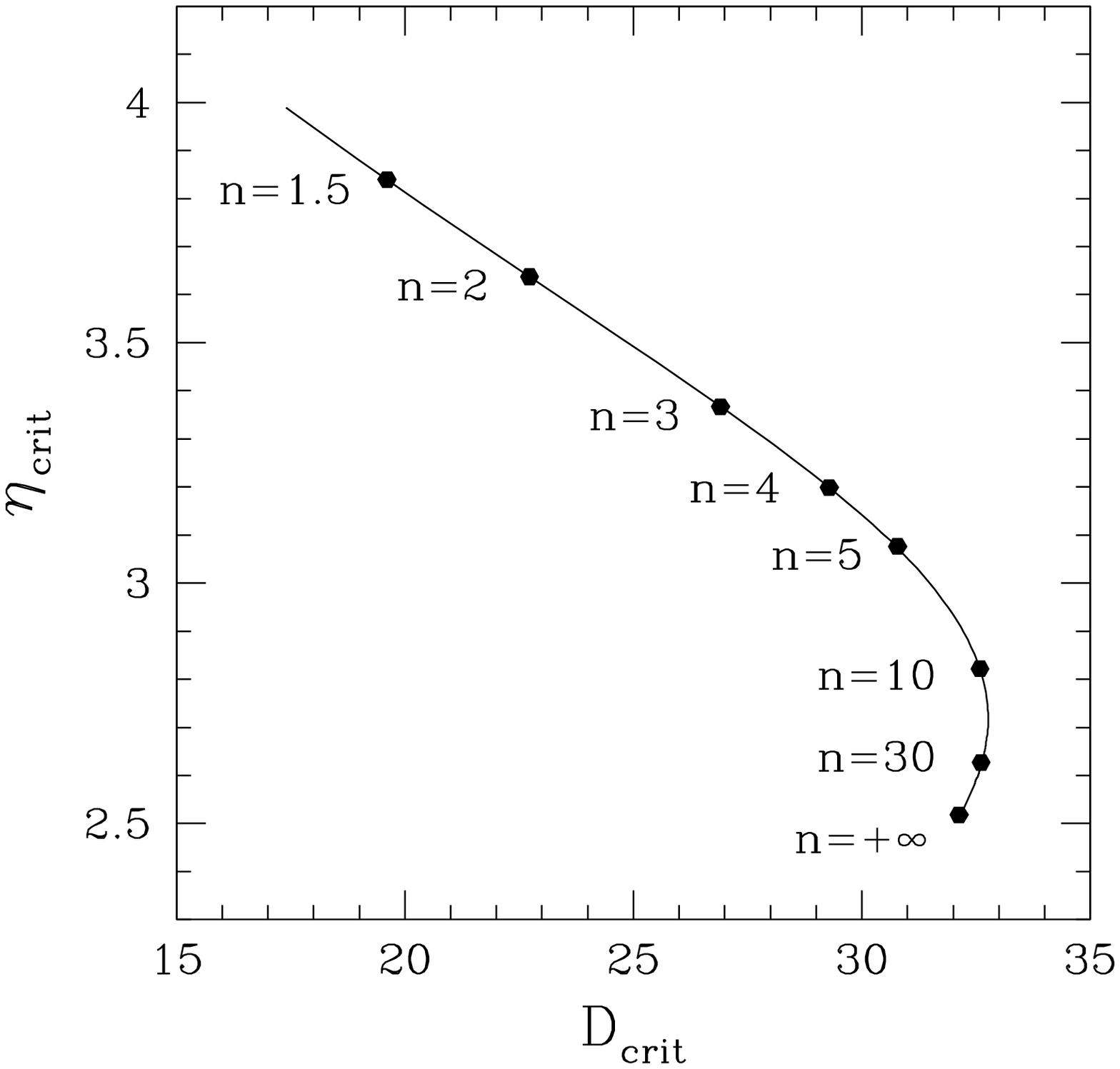}
    \caption{Critical values $(D_{\rm crit}, \eta_{\rm crit})$ 
      for a system in contact with a thermal bath. }
    \label{fig: eta_crit}
  \end{center}
\end{figure}

%
%
%
%
%
In this paper, 
we revisited the issues on the thermodynamic properties of 
stellar self-gravitating system arising from the Tsallis entropy, 
with a particular emphasis to the standard framework using 
the normalized $q$-values. It then turns out that 
the new extremum-entropy state essentially remains unchanged from the 
previous study and is characterized by the 
stellar polytrope, although the distribution function shows several 
distinct properties. Taking these facts carefully, 
the thermodynamic temperature of the extremum state was identified 
through the modified Clausius relation and the specific heat was 
evaluated explicitly. The detailed discussion on the behavior of 
specific heat finally leads to the conclusion that the stability of the 
system surrounded by a thermal wall(canonical case) is drastically 
changed from previous result, 
while the onset of gravothermal instability remains unchanged 
for a system confined in an adiabatic wall(micro-canonical case).  
The existence of these thermodynamic instabilities can also be deduced 
from the variation of entropy and free-energy rigorously.  
As a result, above the certain critical values 
of $\lambda$ or $D$, 
the thermodynamic instability appears at $n>5$ for a system confined in 
an adiabatic wall. As for the system in contact with a thermal bath, 
the onset of thermodynamic instability appears at 
$(\eta_{\rm crit},D_{\rm crit})$ with any value of $n$. In 
Figs. \ref{fig: lambda_crit} and \ref{fig: eta_crit}, 
for the sake of the completeness, the 
critical values evaluated from the conditions 
(\ref{eq: marginal_stability(1)}) and (\ref{eq: marginal_stability(2)}) 
are summarized, respectively. Note also the fact that the critical values in 
Fig. \ref{fig: lambda_crit} exactly coincide with the previous results 
(c.f. Table 1 in paper I).

The one noticeable point in the present result is that the 
macroscopic relation such as the radius-mass-temperature relation 
or the specific heat as a function of density contrast can be 
solely determined from the Emden solutions without any uncertainty. 
Indeed, previous study using the standard linear means 
seriously suffer from the residual dimensional parameter 
$h=(l_0v_0)^3$, which must be practically disappeared from the 
macroscopic description. In paper II, 
the origin of this residual dependence has been addressed in 
connection with the non-extensivity of the entropy. Although the 
scaling relation appeared in the old results is simply 
deduced from the asymptotic behavior of the Emden solutions, 
the explicit $h$-dependence itself has originated from the 
radius-mass-temperature relation. By contrast, in present case,  
the radius-mass-temperature relation was 
derived from the non-trivial relation (\ref{eq: beta-P_relation}), 
in which no such $h$-dependence appears. 
The resultant specific heat is thus obtained free from the residual 
dependence, which is a natural outcome of the 
new framework using the normalized $q$-values. Therefore,  
it seems likely that the new formalism provides a better 
characterization for non-extensive meta-equilibrium state.

In fact, for several application of the new formalism, 
the equation of state for non-extensive system has been 
found to be similar to the result from the ordinary extensive 
thermodynamics. For instance, the equation of state for 
the classical gas model without interaction reduces to the 
ideal gas sate even in the power-law nature of velocity 
distribution \cite{AMPP2001}. Indeed, this fact can be clearly seen 
in our case neglecting the gravity. Recalling that 
both the density and the pressure become {\it homogeneous} 
in the limit $G\to0$,  the equation (\ref{eq: Tphys_P}) immediately 
reduces to the relation $\Tphys=PV$, i.e., $P\propto \rho\Tphys $, 
where $V$ denotes the volume of the system. 
This result apparently seems contradiction with the polytropic relation 
(\ref{eq: polytrope}), however, 
it turns out that the dimensional 
constant $K_n$ immediately yields the relation $K_n=P/\rho^{1+1/n}$ 
from the definition (\ref{eq: K_n}) and thereby 
the polytropic relation makes no sense in the limit $G\to0$. 
Thus, the stellar polytropic system using normalized $q$-values 
successfully recovers the ideal gas limit, in contrast to the old results 
(see Appendix B in paper II). In other words, if the interaction is turned 
on, the equation of state for the ideal gas is no longer valid and 
instead the polytropic relation holds. 
In Sec.\ref{sec: polytrope},  the polytropic relation was 
derived without assuming any specific choice of the gravitational 
potential. This in turn suggests that at least in the mean-field treatment, 
the polytropic relation is a common feature in presence of long-range 
interaction and it generally holds for non-extensive quasi-equilibrium system.

Finally, we have firmly confirmed that the stellar 
polytropic system can be consistently characterized as a plausible 
meta-equilibrium state in the new framework of the non-extensive 
thermostatistics, as well as in the old formalism. At present, however, 
one cannot rigorously discriminate the correct and the applicable formalism 
among them (there might be another possibility that both the formalism become 
indeed correct depending on the situations). While the new framework 
by means of the normalized $q$-values has theoretically desirable 
properties as mentioned above, the results in old formalism might 
provide an interesting connection with dynamical instability of 
gaseous system, as has been suggested by Chavanis 
\cite{Chavanis2002a,Chavanis2002b}. In paper II, 
the thermodynamic instability of stellar polytropic system in the canonical 
ensemble case is shown to exist at the polytrope indices $n>3$, which 
is exactly the same condition as derived from the gaseous system.  
On the other hand, in the new formalism, there is no such 
correspondence, since the instability is completely different from 
that of the gaseous system (see Sec.\ref{subsec: free-energy}). 
This means that the choice of the statistical average is crucial for 
the thermodynamic properties and the physical reason for this discrepancy 
should be further clarified. In any case, in order to pursue 
the physical reality of the non-extensive thermostatistics, 
the limitation of thermodynamical approach is now apparent and the detailed 
kinematical study based on the Boltzmann or the Fokker-Planck equation 
provides an deep insight into the thermal transport property. 
In the light of this, the long-term stellar dynamical evolution by $N$-body 
simulation also becomes useful and the analysis is now in progress. 
The results will be presented elsewhere.

\bigskip
We are grateful to Prof. S. Abe for invaluable discussions and comments. 
This work is supported in part by the Grant-in-Aid for Scientific Research 
of Japan Society for the Promotion of Science (No.$1470157$). 
\clearpage
%
%
%
%
%
\section*{Appendix A: Extremum-state of Tsallis entropy from the OLM method}
\label{appen_A}
%
%
%
%
%
Here, owing to the maximum Tsallis entropy principle, 
we derive an extremum state of the stellar quasi-equilibrium distribution 
by means of the OLM method. According to Ref.\cite{MNPP2000}, 
with a help of the relation (\ref{eq: escort_dist}), 
the variational problem that extremizes the entropy under the 
energy constraint becomes 
\begin{equation}
  \delta\left[S_q\,-\alpha\left\{\int d^6\bftau\,\,p-1\right\}
-\beta \,\int d^6\bftau\left\{ M\left(
\frac{1}{2}v^2+\frac{1}{2}\Phi\right)-E\right\}\,p^q\,
\right]=0.
\end{equation}
Then the variation with respect to probability function $p(\xx,\vv)$ 
leads to 
\begin{eqnarray}
&&  \int d^6\bftau\left[-\frac{1}{q-1}(q\,p^{q-1}-1)\,
    -\alpha-\beta \,M\left( 
\frac{1}{2}\,v^2-\frac{E}{M}\right)\,q\,p^{q-1}\,\right]\delta p
\nonumber
\\
&& ~~~~~~~~~~~~~~~~~~~~~~~~~~~~~~~~~~~~~~~~~~
-\frac{1}{2}\,\beta\,
\delta\left(\Zq\,\int d^6\bftau\,\Phi\,f\right)=0.
\label{appendix_A: variation_OLM_1}
\end{eqnarray}
In the above equation, the last term in the left-hand-side is 
rewritten with 
\begin{eqnarray}
\delta\left( \Zq\int d^6\bftau \,\Phi\, f\right)&=&
\delta\Zq
 \int d^6\bftau\,\Phi\,f+2\,\Zq\int d^6\bftau\,\Phi\delta f
\nonumber \\
&=& q\int d^6\bftau\left\{2M\,\Phi-\int d^6\bftau'\,\Phi(\xx')
f(\xx',\vv')\right\}p^{q-1}\,\delta p, 
\label{appendix_A: variation_phi_f}
\end{eqnarray}
with a help of the equation (\ref{eq: delta_f}). In the last line, 
we have exchanged the role of the variables 
$(\xx,\vv)\leftrightarrow(\xx',\vv')$.  Substituting 
(\ref{appendix_A: variation_phi_f})  into 
(\ref{appendix_A: variation_OLM_1}), we arrive at 
\begin{eqnarray}
  \int d^6\bftau \left[-\frac{1}{q-1}\left(q\,p^{q-1}-1\right)
-\alpha-\beta \,M\,q\,p^{q-1}\left(\frac{1}{2}\,v^2+\Phi-\varepsilon 
\right)\right]\delta p = 0, 
\nonumber
\end{eqnarray}
where the quantity $\varepsilon$ is given by (\ref{eq: epsilon}). 
Recalling the fact that the above equation holds independently of the 
choice of the variation  $\delta p$, we obtain 
\begin{eqnarray}
-\frac{1}{q-1}\left(q\,p^{q-1}-1\right)
-\alpha-\beta \,M\,q\,p^{q-1}\left(\frac{1}{2}\,v^2+\Phi-\varepsilon 
\right)=0, 
\nonumber
\end{eqnarray}
which leads to the power-law distribution. The resultant 
expressions for the one-particle distribution function 
(escort distribution) is summarized in 
equations (\ref{eq: extremum_state}) and (\ref{eq: A_Phi0}), 
with the identification $\Zq'=1$.  
%
%
%
%
%
\section*{Appendix B: Stellar polytropic system characterized by 
        Emden solutions}
\label{appen_B}
%
%
%
%
%
%
%
%
%
In this appendix, we briefly describe the equilibrium 
configuration of stellar polytrope in the spherically symmetric case,   
together with some useful formulae which has been used in 
the main analysis of 
section \ref{sec: thermodynamics} and \ref{sec: stability_variation}.

First notice that the one-particle distribution function 
(\ref{eq: extremum_state}) does not yet completely specify 
the equilibrium configuration, 
due to the presence of gravitational potential which implicitly 
depends on the distribution function itself. Hence, we need to 
specify the gravitational potential or density profile.
>From the gravitational potential (\ref{eq: def_Phi}), it reads 
\begin{equation}
\label{eq: poisson_eq}  
 \frac{1}{r^2}\frac{d}{dr}\left\{r^2\frac{d\Phi(r)}{dr}\right\}=
  4\pi G \rho(r).
\end{equation}
Combining the above equation with (\ref{eq: density}), we 
obtain the ordinary differential equation for $\Phi$. 
Alternatively, a set of equations which represent 
the hydrostatic equilibrium are derived using 
(\ref{eq: poisson_eq}), (\ref{eq: density}) and (\ref{eq: pressure}):  
\begin{eqnarray}
& \frac{dP(r)}{dr}\,=&\,\,-\frac{Gm(r)}{r^2}\,\rho(r), 
\label{eq: hydro_1}
\\
& \frac{dm(r)}{dr}\,=&\,\,4\pi\rho(r)\,r^2.  
\label{eq: hydro_2}
\end{eqnarray}
The quantity $m(r)$ denotes the mass evaluated at the radius $r$ 
inside the wall. Denoting the central density and pressure by 
$\rho_c$ and $P_c$, we then introduce the dimensionless quantities: 
\begin{equation}
\label{eq: dimensionless}
 \rho=\rho_c\,\left[\theta(\xi)\right]^n,\,\,\,\,\,\,
r=\left\{\frac{(n+1)P_c}{4\pi G\rho_c^2}\right\}^{1/2}\,\xi, 
\end{equation}
which yields the following ordinary differential equation: 
\begin{equation}
 \theta''+\frac{2}{\xi}\theta'+\theta^n=0,
\label{eq: Lane-emden_eq}
\end{equation}
where prime denotes the derivative with respect to $\xi$. 
The quantities $\rho_c$ and $P_c$ in (\ref{eq: dimensionless}) 
are the density and the pressure at $r=0$, respectively. 
To obtain the physically relevant solution of (\ref{eq: Lane-emden_eq}), 
we put the following boundary condition:
\begin{equation}
 \theta(0)=1, \,\,\,\,\,\,\,\theta'(0)=0.    
\label{eq: boundary}
\end{equation}
A family of solutions satisfying (\ref{eq: boundary}) is referred to 
as the {\it Emden solution}, which is well-known in the subject of 
stellar structure (e.g., see Chap.IV of Ref.\cite{Chandra1939}).

To characterize the equilibrium properties of Emden solutions,  
it is convenient to introduce the following 
set of variables, referred to as homology invariants 
\cite{Chandra1939,KW1990}: 
\begin{eqnarray}
 u &\equiv& \frac{d\ln m(r)}{d\ln r}=
\frac{4\pi r^3\rho(r)}{m(r)}=-\frac{\xi\theta^n}{\theta'},
\label{eq: def_u}
\\
\nonumber\\
 v &\equiv&  - \frac{d\ln P(r)}{d\ln r}=
\frac{\rho(r)}{P(r)}\,\,\frac{Gm(r)}{r}
=-(n+1)\frac{\xi\theta'}{\theta},  
\label{eq: def_v}
\end{eqnarray}
which reduce the degree of equation (\ref{eq: Lane-emden_eq}) 
from two to one. The derivative of these variables with 
respect to $\xi$ becomes
\begin{equation}
\frac{du}{d\xi} = \left(3-u-\frac{n}{n+1}\,v \right)\,\frac{u}{\xi},
~~~~~~~~
\frac{dv}{d\xi} = \left(-1+u+\frac{1}{n+1}\,v \right)\,\frac{v}{\xi}.
\label{eq: d(u,v)/dxi}
\end{equation}
Equations (\ref{eq: Lane-emden_eq}) can thus be re-written with 
\begin{equation}
 \label{eq: uv_eqn}
  \frac{u}{v}\,\frac{dv}{du}=\frac{(n+1)(u-1)+v}{(n+1)(3-u)-nv}.   
\end{equation}
The corresponding boundary condition to (\ref{eq: boundary}) 
becomes  $(u,v)=(3,0)$.

Now, utilizing the above homology invariants, 
we list some useful formulae which can be derived 
from the hydrostatic equations (\ref{eq: hydro_1}) and (\ref{eq: hydro_2}) 
(see also Appendix A in paper I): 
\begin{eqnarray}
\int_0^{r_e}dr \,4\pi r^2\,P(r)  &=& - \frac{1}{n-5}
\left\{8\pi\,r_e^3P_e-(n+1)\frac{MP_e}{\rho_e}+\frac{GM^2}{r_e}\right\}
\nonumber \\
&=& -\frac{1}{n-5}\,\,\frac{GM^2}{\re}\,\,
\left(2\,\frac{\ue}{\ve}-\frac{n+1}{\ve}+1\right),
\label{eq: formula_1}
\end{eqnarray}
\begin{eqnarray}
\int_0^{r_e} dr\,\frac{Gm(r)}{r^2}\,4\pi r^3 \rho(r) 
&=& -4\pi r_e^3 P_e + 3 \int_0^{r_e}dr \,4\pi r^2\,P(r), 
\nonumber \\
&=& -\frac{n+1}{n-5}\,\,\frac{GM^2}{\re}\,\,
\left\{ \frac{\ue}{\ve}-\frac{3}{\ve}+\frac{3}{n+1}\,\right\},
\label{eq: formula_2}
\end{eqnarray}
\begin{eqnarray}
\int_0^{r_e}  dr\,\frac{G\,m^2(r)}{r^2} 
&=& -\frac{GM^2}{r_e} \, - \,8\pi r_e^3P_e \,+ \,6 \int_0^{r_e}dr \,
4\pi r^2\,P(r),
\nonumber \\
&=& -\frac{n+1}{n-5}\,\,\frac{GM^2}{\re}\,\,
\left\{ 2\left(\frac{\ue}{\ve}-\frac{3}{\ve}\right)+1\,\right\}.
\label{eq: formula_3}
\end{eqnarray}

Finally, using these formulae, we evaluate the total energy of the 
stellar system in terms of the variables at the edge $\re$: 
\begin{eqnarray}
  E= K+U &=& \frac{3}{2}\,\int_{0}^{\re}dr\,4\pi r^2\,P(r) 
  -\int_0^{\re}\,dt\,\frac{Gm(r)}{r}\,\frac{dm}{dr}
\nonumber \\
&=& -\frac{1}{n-5}\,
\left[\,\frac{3}{2}\left\{ \frac{GM^2}{\re}-(n+1)\frac{MP_e}{\rho_e}\right\}
+(n-2)\,4\pi\,\re^3\,P_e\,\right], 
\nonumber 
\end{eqnarray}
which can be re-expressed in terms of the homology invariants: 
\begin{eqnarray}
E &=& \frac{1}{n-5}\,\frac{GM^2}{\re}\,\,
\left[\,\frac{3}{2}\left\{1-\frac{n+1}{\ve}\right\}+(n-2)\frac{\ue}{\ve}
\,\right].
  \label{eq: energy_uv}
\end{eqnarray}
%
%
%
%
%
%
%
%
%
\clearpage
%
%
%
%
%
%

%
%
%
%
%
%
%
%
%
%
%
%
%
%
%
%
%
%
%
%
\clearpage
%
%
%
%
%
%
%
%
%
%
%
%
%
%
%
%
%
%
%
%
%
%
%
%
%
%
%
%
%
%
%
%
%
%
%
%
%
%
%
%
%
%
%
%
%
%
%
\end{document}